\documentclass[preprint,12pt]{elsarticle}

\usepackage{amsmath,color,soul}
\soulregister\cite{7}
\soulregister\ref{7}
\soulregister\eqref{7}
\usepackage[colorlinks=true, citecolor=blue]{hyperref}%
\journal{Powder Technology}

\begin{document}

\begin{frontmatter}

\title{On the Effect of Stress Dependent Interparticle Friction in Direct Shear Tests}

\author[label1]{Bettina Suhr\corref{cor}}
\cortext[cor]{Corresponding author}
\author[label1]{Klaus Six }
\address[label1]{Virtual Vehicle Research Center, Inffeldgasse 21/A, A-8010 Graz, Austria, bettina.suhr@v2c2.at, klaus.six@v2c2.at,  http://www.v2c2.at}

\begin{abstract}
Contact friction is a key influence factor for the shearing behaviour of granular media. In the Discrete Element Method (DEM)   contact friction is usually modelled with Coulomb's law assuming a  constant interparticle friction coefficient. From tribology it is known that friction is influenced by several factors, e.g.,~temperature, normal stress, surface condition, etc. None of these influences can be modeled with the constant interparticle friction coefficient from Coulomb's law.  
 For a given granular material (particle shape distribution), the usage of  constant interparticle friction  in DEM models generally results in constant bulk friction coefficients in the simulation of direct shear tests. 
 While this is frequently seen in experiments with equi-sized spherical particles, papers exist in literature  which report a stress dependency of bulk friction for non-spherical particles of certain materials. 
 In this work, a stress dependency of bulk friction is obtained by introducing a model for stress dependent interparticle friction in DEM simulations. 
 For validation experimental results of direct shear tests conducted on single or paired glass beads are used. 
 While the bulk friction of paired spheres clearly decreases with increasing normal stress, it is nearly constant for single spheres. 
 DEM simulations with the stress dependent interparticle friction  are in good accordance with the experimental results of both single and paired spheres.
  A comparison with simulations, using constant interparticle friction, clearly shows the advantage of the proposed model.
  
 \end{abstract}


\end{frontmatter}

\section{Introduction}

In DEM simulations of granular material, the modelling of friction at particle-particle and particle-environment contacts has a significant influence on the predicted shear behaviour of the bulk material. In the sense of a tribological system friction is influenced by several parameters like contact normal load, relative motion, surface roughness, contact temperature and contact conditions (dry, wet, lubricated, \ldots), etc.

Regarding the simulation of the mechanical behaviour of solid-like granular materials, the discrete (distinct) element method (DEM), as introduced by Cundall and Strack, \cite{Cundall1979}, is a widely used tool. 
In DEM the force, which results at each contact,  is  decomposed in normal and tangential direction, $F_n$ and $F_t$. Several models exist for the calculation of both quantities. 
 In tangential direction, the force, which can be transferred, is bounded. The commonly used model is Coulomb's law with a constant interparticle friction coefficient, $\mu$. 
  For cohesionless materials Coulomb's law can be written as follows:   
\vskip-.6cm
\begin{eqnarray}
F_t=\min\left( \mu F_n, \tilde{F_t}\right)\; ,
\end{eqnarray}
where $ \tilde{F_t}$ is the pre-sliding shear force calculated using the contact constitutive model. 
  Coulomb's law can also be  stated using the  internal friction angle, $\phi$, which is connected to the  interparticle friction coefficient by $\mu=\tan(\phi)$.

 Frequently used tests for the investigation of the shear behaviour of granular materials are the triaxial test and the direct shear (or shear box) test.  
Usually, the Mohr-Coulomb failure criterion is used which reads as:
\vskip-.6cm
\begin{eqnarray}\label{eq:MC}
\tau_f=\tan(\Phi) \sigma_n + c\; ,
\end{eqnarray}
where $\tau_f$ is the final shear stress, $\Phi$ is the  bulk friction angle and $c$ is a material parameter representing cohesion of the granular material, i.e.,~$c=0$ for cohesionless materials. The bulk friction angle of a granular material is an important characteristic for its shear behaviour. Alternatively, the peak friction angle can be  determined, where the maximal shear stress instead of the final one is used in equation \eqref{eq:MC}.

The bulk friction angle of a granular material depends on the porosity of the packing as well as on particle properties, e.g.,~size, shape and surface roughness. Regarding particle roughness several works in literature state a strong influence of the interparticle friction on the bulk friction angle, e.g.,~in \cite{Ni2000, Cui2006, Haertl2008}, direct shear tests were simulated and compared to experiments.

Many papers in the literature state that the bulk/peak friction angle is constant, i.e.,~independent of the normal stress. In the opinion of the authors of this work this has mainly two reasons. One reason is that often  equi-sized spherical particles are considered, where the dependency of the bulk friction coefficient on the normal stress is usually negligible. The second reason is that the way of analysing the results can sometimes be misleading. A frequently found plot is  shear stress over  normal stress. Here it is very hard to see deviations from the linear trend. To investigate a stress dependency of the bulk friction coefficient other representations can be more helpful, e.g.,~bulk friction over normal stress or bulk friction over porosity. This point will be addressed also later on.  

Regarding equi-sized spherical particles one example is the work of Cui and O'Sullivan, \cite{Cui2006}, who conducted direct shear tests on steel balls. Within the regime of applied normal stresses (55~kPa - 164~kPa), a linear relation between the measured shear stress and normal stress was found. This justifies the application of the Mohr-Coulomb criterion, and the bulk friction angle can be considered constant. 

H\"artl and Ooi carried out direct shear tests on single and paired glass beads, see \cite{Haertl2008} and \cite{Haertl2011}.  The applied normal stress  ranged from 3~kPa to 24~kPa. %
 For the single glass beads the relation between shear stress and applied normal stress was nearly linear, thus the bulk friction angle was constant. 
 On the contrary, a clearly non-linear relation between shear stress and applied normal stress was found for the paired glass beads. This nonlinearity was hardly seen in the plot of shear stress over normal stress. However, when the bulk friction coefficient over initial porosity was plotted, the stress dependency of the bulk friction angle was clearly seen.  %
  
Indraratna el~al.~found similar experimental results in \cite{Indraratna2011}, regarding a stress dependency of the bulk friction angle of railway ballast  in direct shear tests. The normal stress was varied  between 15~kPa and 75~kPa and a nonlinear dependency between shear stress  and normal stress was shown.  In several citations of works on rock-fill materials, given in this work, a non-linear relationship is stated, which is significant at low normal stresses and gradually reduces as the normal stress increases.  

This description matches well with the results of Tuzun and Walton, see \cite{Tuzun1992}. A stress dependent coefficient of friction between smooth silo walls and particles was found for small normal stresses.  It seems that depending on the considered material and particle shape, a non-linear relation between shear stress and normal stress can be observed for low normal stresses. 

The above experimental findings on granular materials and results obtained by the authors on the wheel-rail contact for steel, were the  motivation   to introduce  a non-constant coefficient of friction in DEM simulations. This aims at an improved prediction of the observed normal stress dependency of the bulk friction angle. %
 In \cite{Suhr2015},  Suhr and Six,  conducted DEM simulations   using the contact model with normal stress dependent friction. Direct shear tests  on steel spheres of a given size distribution were considered. Steel was chosen as material for the spheres, as the experiments for the stress dependency of the interparticle friction coefficient were given for steel, see \cite{Popov2002}.  
  Although no experimental results on the direct shear tests were available for a quantitative comparison, the following qualitative behaviour of the granular material was found. A variation of the constant interparticle friction coefficient demonstrated a strong dependency of the bulk friction angle on interparticle friction. For a  constant interparticle friction coefficient the resulting bulk friction angle showed no normal stress dependency. When the non-constant (stress dependent) friction coefficient was used, the stress dependency of the bulk friction angle could be seen clearly.

In this paper, the contact model with normal stress dependent friction will be used to investigate the normal stress dependency of the bulk friction coefficient, as seen from the experiments on glass beads in \cite{Haertl2008} and \cite{Haertl2011}. 

The paper is organised as follows. In Section 2 an overview of the main experimental findings from the direct shear tests of \cite{Haertl2008} and \cite{Haertl2011} will be given. The third section summarises the used DEM contact model with the stress dependent friction coefficient. In Section 4 DEM simulations with constant and stress dependent interparticle friction coefficient will be presented. The influence of the model parameters will be pointed out and a comparison with experimental results is shown. Finally conclusions are drawn in Section 5. 

\section{Jenike Shear Tests -- Experimental Results}
\begin{figure}
\centering
\parbox{0.55\textwidth}{
\includegraphics[height=5.5cm]{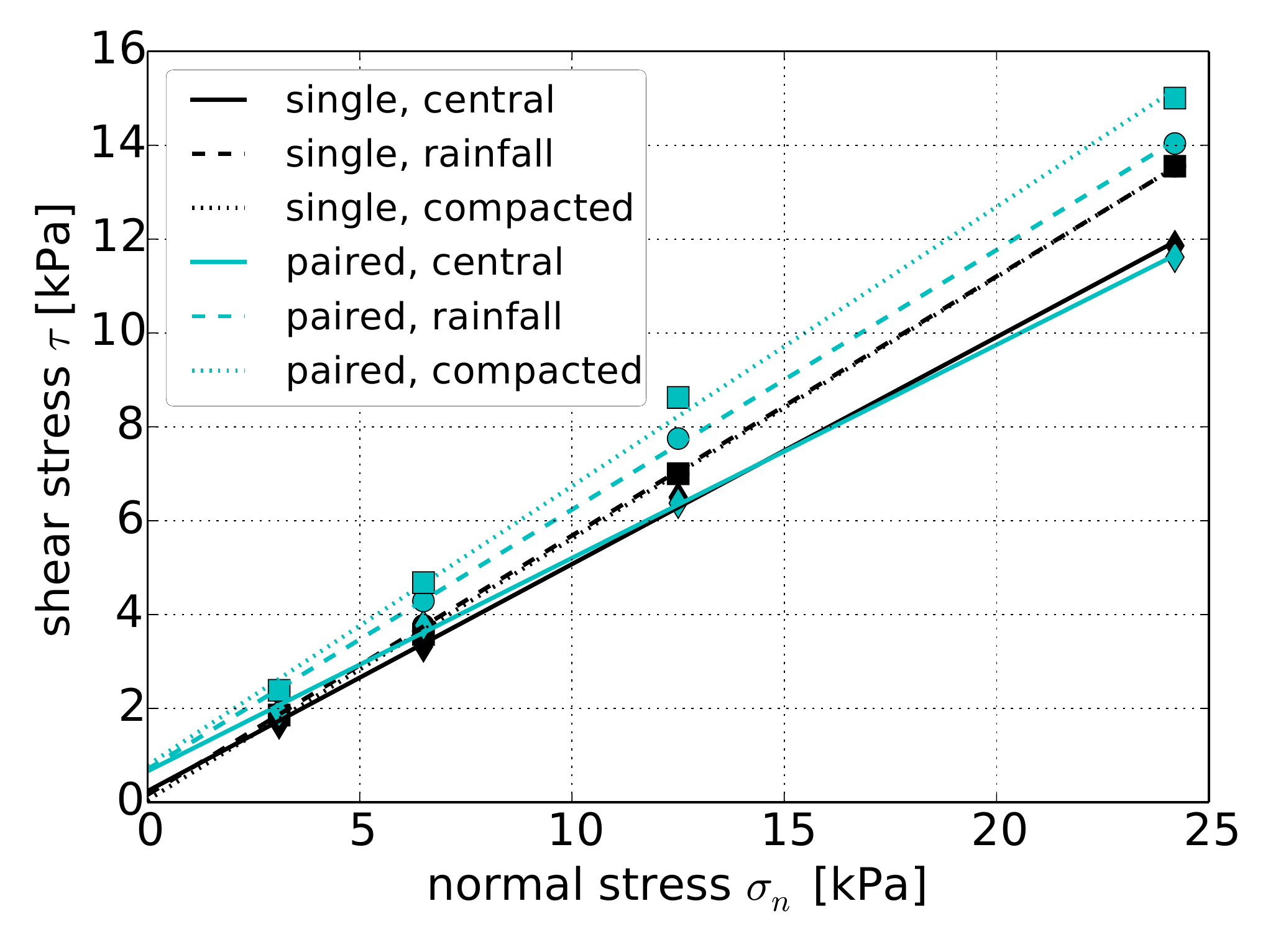}
\caption{Shear stress over applied normal stress of single and paired glass spheres, data taken from \cite{Haertl2008}.}\label{fig:HOclassic}
}
\end{figure}

\begin{figure}
\centering
\parbox{0.55\textwidth}{
\includegraphics[height=5.5cm]{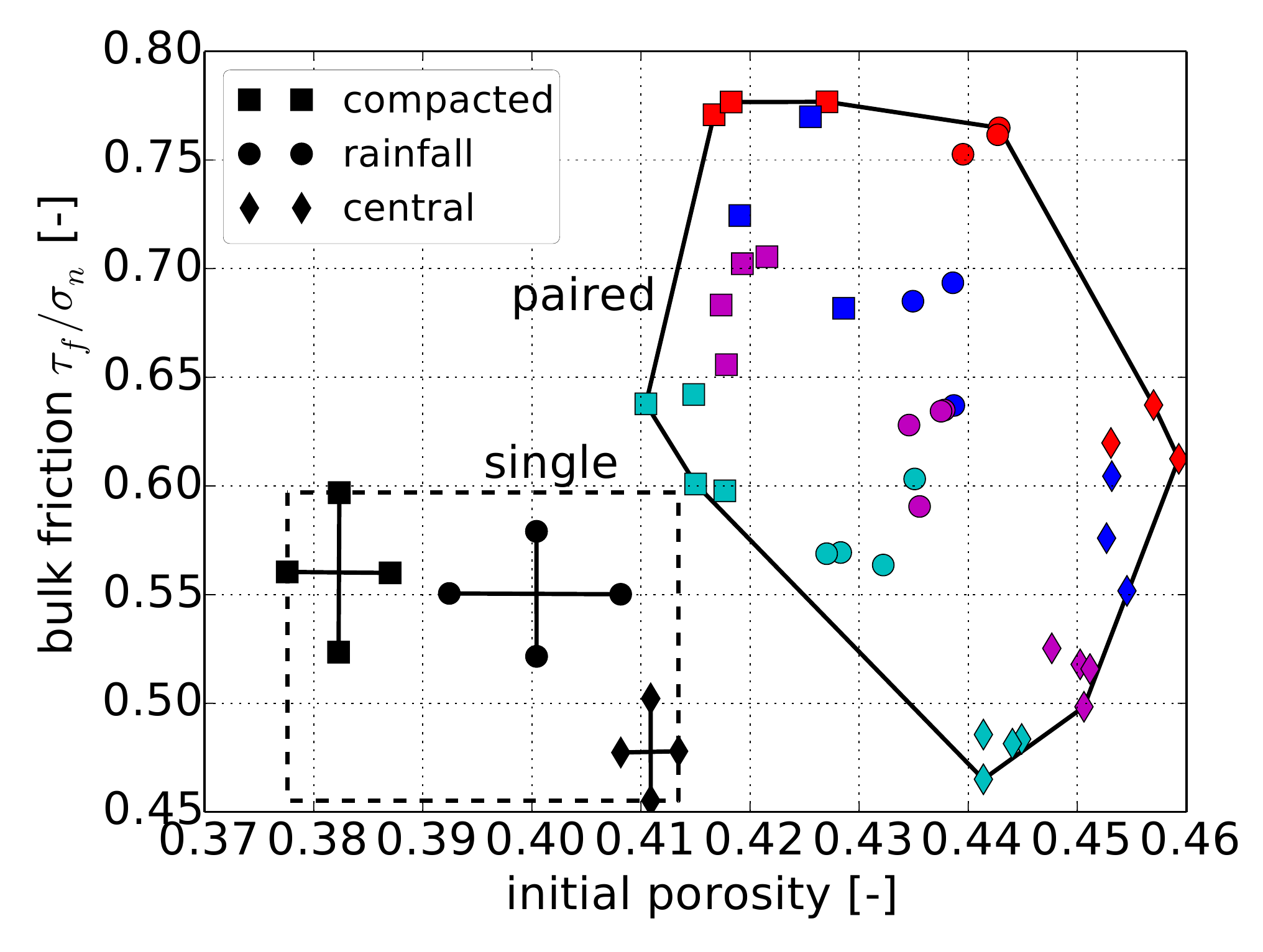}
\caption{Stress dependency in bulk friction, taken from  \cite{Haertl2011}. Marker types corresponding to filling of paired spheres; compacted: square, rainfall: circle, central: diamond. Colouring of applied normal stresses: 3.1~kPa: red, 6.4~kPa: blue, 12.5~kPa: magenta, 24.2~kPa: cyan.}
\label{fig:HOporo}
}
\end{figure}

\begin{figure}
\centering
\parbox{0.55\textwidth}{
\includegraphics[height=5.5cm]{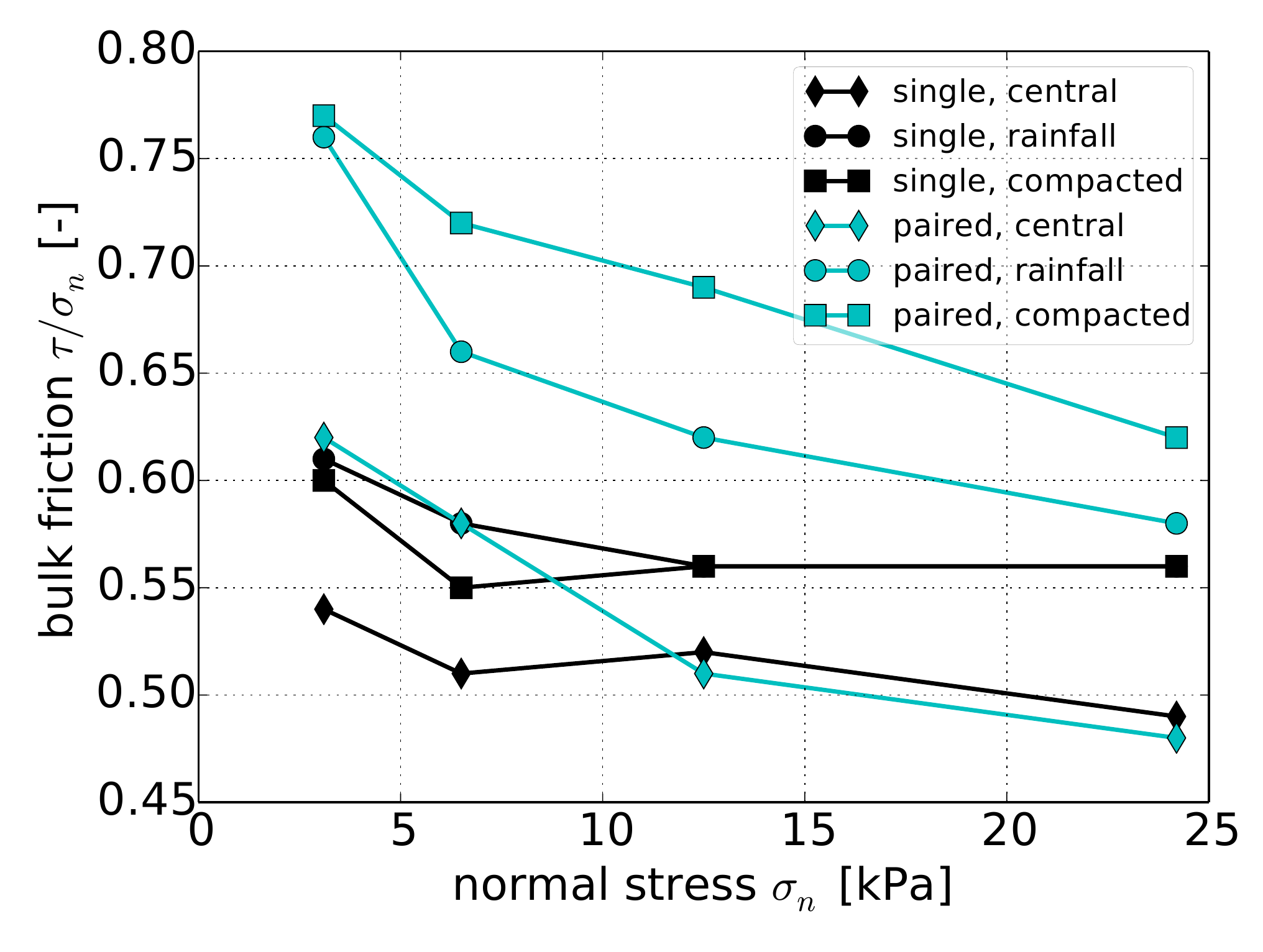}
\caption{Bulk friction coefficient over applied normal stress of single and paired glass spheres, data taken from \cite{Haertl2008}.}\label{fig:HObulkFrict}
}
\end{figure}

All experimental results shown here stem from H\"artl and Ooi, \cite{Haertl2008} and \cite{Haertl2011}. Only the main findings will be shown, for details the reader is referred to the original works. A Jenike shear tester with a cylindrical cell of the diameter 143~mm was used. The height of the lower and upper ring was 19~mm and 24~mm respectively. Both top and bottom plate were roughened with grooves as recommended by the standard, \cite{Schulze1989}. The used glass beads had a diameter of 6~mm. Their material parameters for single spheres can be found in Table~\ref{tab:mat-par}. Paired spheres were made by gluing two spheres together. 
\begin{table}
\begin{tabular}{l l }
 \hline%
 Diameter [mm]& 6\\
 Density [kg/m$^3$]& 2550\\
 Poissons ratio [-]& 0.22\\
 Shear modulus [N/m$^2$]& 1.67e+10 \\
  Friction coefficient (glass-steel)  [-]& 0.13\\
 Friction coefficient (glass-glass)  [-]& 0.2\\
 Coefficient of restitution  [-]& 0.87 \\
\hline%
\end{tabular}
\caption{Material parameters of single glass beads.}\label{tab:mat-par}
\end{table}  
Three different filling methods were compared: central filling (filling through a central funnel), rainfall (filling through a sieve) and compacted filling (manually compaction). The filling methods resulted in different initial porosities. Four different levels of  applied normal stress were considered: 3.1~kPa, 6.4~kPa, 12.5~kPa and 24.2~kPa. For single as well as for paired spheres for each filling method and each level of applied normal stress three repetitions of the experiment were conducted. The classical plot of shear stress over applied normal stress is shown in Figure~\ref{fig:HOclassic}. The least square fit through the data is also presented. In this representation it is hard to notice the normal stress dependency of the shear stress. This is different in the next Figure~\ref{fig:HOporo}, where the bulk friction is plotted over the initial porosity. This Figure is obtained  by plot digitalisation and thus may contain small inaccuracies. It can be seen that the single spheres have a lower porosity than the paired spheres. The filling method has a strong influence on the initial porosity, where the central filling gives the loosest packings and the compacted filling gives the densest packings. The lower the initial porosity the higher is the resulting bulk friction coefficient  for both single and paired spheres. The results of the single spheres show no clear dependency on the normal stress. On the contrary, the influence of the normal stress on the bulk friction coefficient of the paired spheres is evident.  
In Figure~\ref{fig:HObulkFrict} the averaged bulk friction coefficient for each level of normal stress is displayed, for single and paired spheres separately. In this type of plot it is easier to quantify the extend of stress dependency.

\section{Stress Dependent Interparticle Friction in DEM}
\subsection{Derivation of the Stress Dependent Interparticle Friction Model}

\begin{figure}
\includegraphics[width=0.48\textwidth]{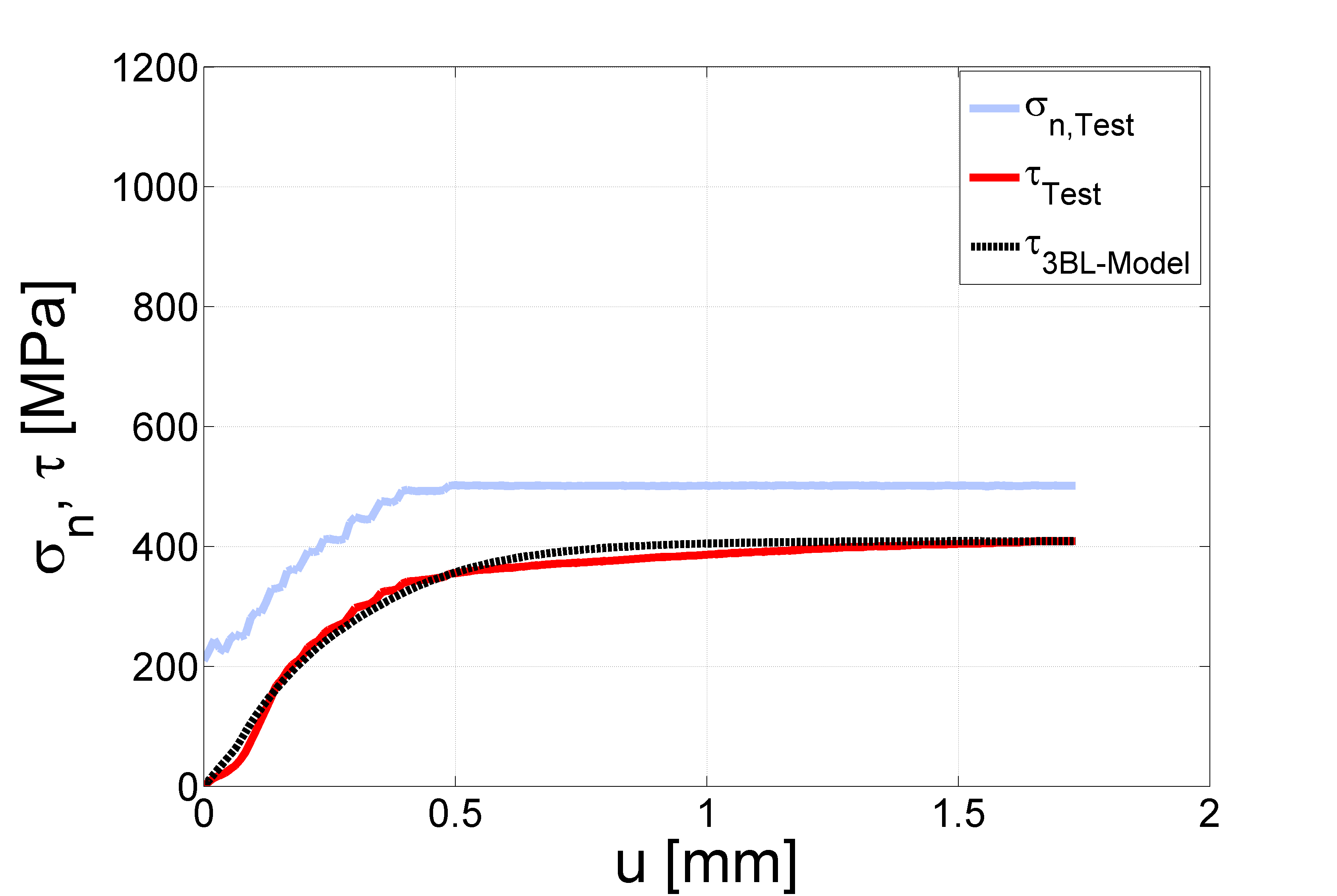}%
\hfill
\includegraphics[width=0.48\textwidth]{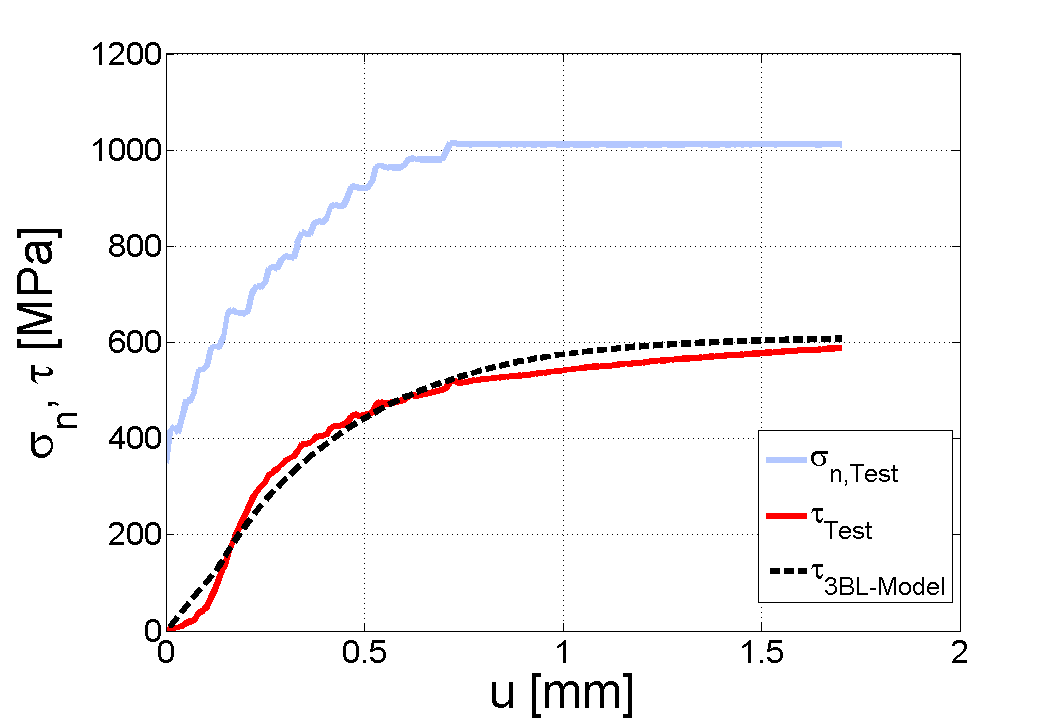}%
\caption{High Pressure Torsion (HPT) tests measurement of normal stress ($\sigma_n$) and shear stress ($\tau$) over displacement $u$. Left:  $\sigma_n$ = 500~MPa. Right: $\sigma_n$ = 1000~MPa. Increasing normal stress reduces ratio $\tau$/$\sigma_n$. Comparison to results from model \cite{Six2015}.}\label{fig:HPT}
\end{figure}

In tribology it is known that Coulomb’s law, with its one constant friction coefficient, is not sufficient to model frictional contacts under certain conditions. As many experiments show, applied normal stress, sliding velocity, temperature, surface conditions as well as other factors can severely influence the observed frictional behavior. A general description of these phenomena can be found e.g.~in \cite{Mezenes2013}. The normal stress dependency of the friction coefficient is reported in the literature for several materials e.g.~for  aluminum or polymers, see e.g.~\cite{Chowdhury2011, Quaglini2011}, and for E-glass fiber reinforced epoxy composites, see \cite{El-Tayeb1996}.   
For the authors of the current paper, the introduction of a pressure dependent friction coefficient was initially motivated by works on wheel-rail contact (steel-steel). In \cite{Six2015}, Six et al.,~conducted High Pressure Torsion tests (HPT). In a HPT test two steel discs are rotated against each other, while the normal stress, $\sigma_n$, and the shear stress, $\tau$, are measured. In this case the ratio between   $\tau$ and $\sigma_n$ is the coefficient of friction. It showed that the assumption of a constant coefficient of friction is not sufficient to reproduce results observed at the experiments, see Figure~\ref{fig:HPT}.  From the left plot to the right plot the maximum normal stress $\sigma_n$ is doubled. Assuming a constant coefficient of friction, then $\frac{\tau}{\sigma_n}$ would be constant,  thus $\tau$ would be doubled. Comparing the value of $\tau$ in the left and right part of  Figure~\ref{fig:HPT}, it can be seen that the $\tau$ is increased not by a factor  2 but 1.5.  Therefore, a significant dependency of the coefficient of friction on the normal load can be concluded from the experiments.

In \cite{Popov2002},  Popov~et~al.,~obtained similar results for steel-steel contacts. The method of Movable Cellular Automata (MCA) was used to model the wheel-rail contact, and from simulations a normal pressure dependent coefficient of friction was derived:
\vskip-.6cm
\begin{eqnarray}\label{eq:pdfPopov}
\mu(\sigma_n)=\mu_0 + \frac{c_1}{1+c_2 \,\sigma_n }\; ,
\end{eqnarray}
where $\sigma_n$ is the applied normal stress and $\mu_0, c_1, c_2$ are model parameters which need to be adapted to the considered material. Popov et al.~specified the parameter values for steel-steel contact, but these values will not be used here. 

   It is future work to obtain experimental data on  the frictional behaviour of glass-glass and glass-steel contacts for the determination of the three model parameters.  Measurements of the coefficient of friction over a wide range of applied normal stresses will be necessary.  In literature only experimental works on glass-glass contacts for few, separate normal stresses can be found. They indicate a normal stress dependency of the friction coefficient. Another problem is that measurements of the friction coefficients for sphere-sphere or sphere-plane contacts, mostly use small spheres. According to Hertz' contact law,  for small spheres already very small normal loads  result in high averaged normal pressures. 
 Thus, it is difficult to measure the friction coefficient for small normal pressures, where the model predicts a steep decrease. 
  In \cite{Cavarretta2010}, Cavaretta et al.,~used small and large glass ballotini (1 - 3~mm diameter). The minimal applied forces resulted in mean normal stresses above 386~MPa for large ballotini and 663~MPa for small ballotini. The measured friction coefficients lied between 0.158 and 0.176. In \cite{Procter1974}, Proctor et al.,~used glass spheres between 3 and 4~mm diameter and stresses lied between 90~MPa and 196~MPa. They reported a scattering in the measured friction coefficient between 0.087 and 0.176. In the same work, results of Tong, \cite{Tong1970}, are cited. Here glass spheres of 3 mm diameter roll over a plane, and the normal force is  caused by gravity. The resulting average stress is as low as  16~MPa, and considerably higher values of the friction coefficients are measured: 0.267 – 0.287. 
  When the difficulties in measuring dry glass-glass contacts for spheres are considered, the overall trend of a decrease in the fiction coefficient with increasing stress is supported by the experimental results.  %

  In this work, it is assumed that  glass-glass and glass-steel contacts generally show a similar behaviour. Based on this assumption the according model parameters are searched to reproduce the observed bulk material behaviour. To avoid an over-parametrized model, only one function is used to average over glass-glass and glass-steel contact. A parameter study will be carried out, to investigate the influence of the parameters on the model output. 

\subsection{Formulation of the DEM Contact Model with Stress Dependent Friction}

In the DEM simulation the simplified Hertz Mindlin contact model without miscroslip and with damping as described by Tsuji et al., \cite{Tsuji1992}, and Antypov et al.,~\cite{Antypov2011}, will be used. 

In normal direction of the contact the Hertz model is applied. Introducing the  the equivalent Young modulus of the contact, $\hat{E}$,  the equivalent contact radius, $\hat{R}$, and  the overlap in normal direction, $u_n$ , the normal force is given as:
\vskip-.6cm
\begin{eqnarray}\label{eq:Hertz}
F_n=\frac43 \hat{E} \sqrt{\hat{R}} \sqrt{u_n^3}\; .
\end{eqnarray}
The averaged normal pressure on a contact, $\bar{\sigma}_n$ can be calculated by dividing the contact force by the contact area.
In the Hertzian contact model the area of contact is circular (sphere-sphere contact), and the contact radius is given by $a = \left( \frac{3F_n \hat{R}}{4\hat{E}}\right)^{\frac13}$. Then, it follows that
\vskip-.6cm
\begin{eqnarray}\label{eq:Hertz_pm}
\bar{\sigma}_n:=\frac{F_n}{a^2 \pi}= \frac{F_n}{\pi} \left( \frac{4\hat{E}}{3F_n \hat{R}}\right)^{\frac23}\; .
\end{eqnarray}
In the tangential direction of the contact the Mindlin  model without microslip is applied. 
In time step $k$ the trial or pre-sliding shear force  is denoted by $F_{t,t}^k$ and is calculated incrementally using the last time step's value, $F_t^{k-1}$:
\vskip-.6cm
\begin{eqnarray}\label{eq:Mindlin_trial}
 F_{t,t}^k = F_t^{k-1} +  \Delta F_{t,t} \qquad \Delta F_{t,t} = 8\, a\, \hat{G} \, \Delta u_s \; , %
\end{eqnarray}
where  $\hat{G}$ is  the equivalent shear modulus and $\Delta u_s$ the increment of the shear displacement. 
For brevity the index of the time step will be dropped from now on. Using the {\it constant coefficient of friction}, the shear force is given by:
\vskip-.6cm
\begin{eqnarray}\label{eq:Mindlin}
F_{t}=\left\{ 
\begin{array}{ll}
 F_{t,t}& \mbox{ if } F_{t,t}\leq \mu F_n\\
 \mu F_n& \mbox{ otherwise}
 \end{array} \right. \; .
\end{eqnarray}
For the use of the {\it pressure dependent friction coefficient}, we now change Equation~\eqref{eq:Mindlin} to:
\vskip-.6cm
\begin{eqnarray}\label{eq:Mindlin_pdf}
F_{t}=\left\{ 
\begin{array}{ll}
 F_{t,t}& \mbox{ if } F_{t,t}\leq \mu(\bar{\sigma}_n) F_n\\
 \mu(\bar{\sigma}_n) F_n& \mbox{ otherwise}
 \end{array} \right. \; ,
\end{eqnarray}
where $\bar{\sigma}_n$ is given by Equation~\eqref{eq:Hertz_pm}  and $\mu(\bar{\sigma}_n)$ by Equation~\eqref{eq:pdfPopov}.

After the computation of normal and tangential force the damping is applied as  described by Tsuji et al., \cite{Tsuji1992}, and Antypov and Elliott,~\cite{Antypov2011}. The coefficient of restitution is assumed to be the same in normal and in tangential direction and is denoted with $\epsilon$. Then, the dissipative forces in normal and tangential direction are given by:
\vskip-.6cm
\begin{subequations}
\begin{align}\label{eq:Damped}
F_{\mbox{dis}}^n=& \eta_n \dot{u}_n & F_{\mbox{dis}}^t=& \eta_t \dot{u}_s\; \\
\eta_n=& \alpha(\epsilon) \sqrt{\hat{m} K} u_n^\frac14 & \eta_t=&\eta_n \; \\
\alpha(\epsilon)=&-\sqrt{5}\frac{\ln(\epsilon)}{\sqrt{\ln^2(\epsilon) + \pi^2}} &
K=&\frac43 \hat{E} \sqrt{\hat{R}} \quad ,
\end{align}
\end{subequations}
 where $\hat{m}$ is the equivalent mass.Thus, the contact force in normal direction is computed as follows:
\vskip-.6cm
\begin{eqnarray}\label{eq:DampedN}
F_n=\frac43 \hat{E} \sqrt{\hat{R}} \sqrt{u_n^3} + F_{\mbox{dis}}^n\; ,
\end{eqnarray}
while in tangential direction, using the constant coefficient of friction, it holds:
\vskip-.6cm
\begin{eqnarray}\label{eq:MindlinDamped}
F_{t}=\left\{ 
\begin{array}{ll}
 F_{t,t} + F_{\mbox{dis}}^t& \mbox{ if } F_{t,t}< \mu F_n\\%+F_{\mbox{dis}}^t
 \mu F_n& \mbox{ otherwise}
 \end{array} \right. \; .
\end{eqnarray}
As suggested by Cundall and Strack, \cite{Cundall1979}, damping in tangential direction is only applied, if the elastic part, $F_{t,t}$, is smaller than $\mu F_n$. 
For the pressure dependent friction coefficient, Equation \eqref{eq:Mindlin_pdf} is modified analogously.

\section{DEM Simulations of the Jenike Shear Tests}
\subsection{General Simulation Setup}

The DEM simulations presented in this work are carried out with the Open-source software Yade, \cite{Yade2010}.  In this software the soft contact approach is used together with explicit discretization in time.  The simplified Hertz-Mindlin no-slip contact model with damping in its classical form as given in Equations (\ref{eq:DampedN}, \ref{eq:MindlinDamped}) or its modification with stress dependent interparticle friction  will  be used. 

\begin{figure}
\centering
\includegraphics[width=0.7\textwidth]{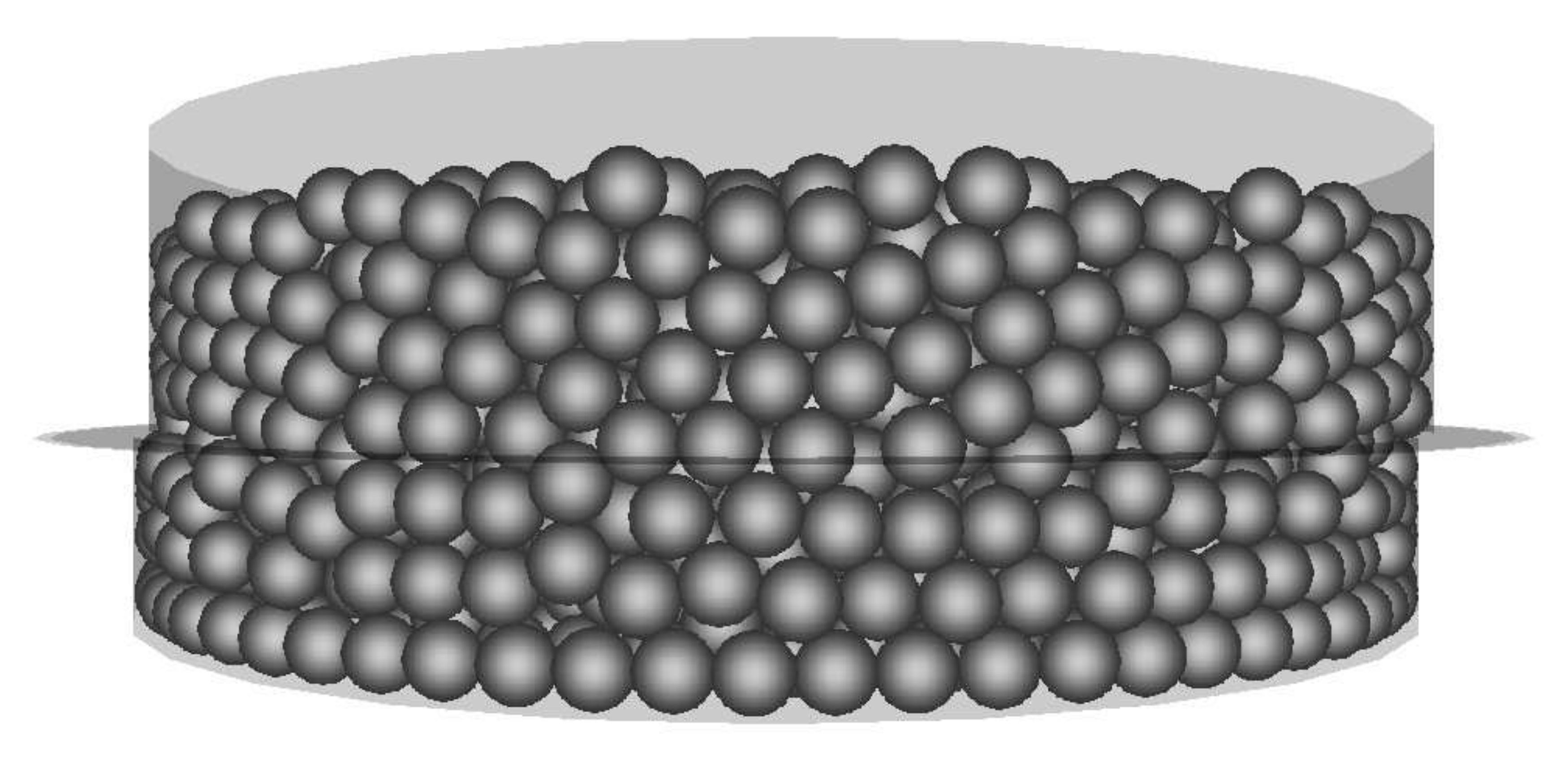}%
\caption{DEM model of Jenike shear cell filled with single spheres.}\label{fig:shearbox}
\end{figure}

The setup of a Jenike shear cell is modeled in the original size. The grooves on the top and bottom plates are neglected for simplicity. The model of the shear cell in its initial position  with a 3~mm offset in lateral direction (according to the stardard for Jenike shear testing, \cite{Schulze1989}) is shown in Figure~\ref{fig:shearbox}. 
 The material parameters for the single glass spheres are given in Table~\ref{tab:mat-par}. The paired spheres are modeled as rigid clumps of two spheres with an aspect ratio of two (i.e.,~no overlap between the spheres). 

For sample generation 5000 single spheres or 2500 clumps of paired spheres  are randomly placed in a box above the shear cell. Then the particles fall under the influence of gravity and are allowed to come to rest in the shear cell. To achieve a dense packing the friction coefficient is set to 0.05 in this initial phase of the simulation. Also, the gravity  is enlarged by  the factor 5. Differences between control samples, which settled under normal gravity, and samples, which settled under increased gravity, were negligible. When the particles  are at rest, all particles which are not completely inside the upper ring are removed. Typical samples of single spheres consisted of 3400 particles and samples of paired spheres consisted of 1630 clumps. A steel plate is inserted above the particles, the friction coefficient  and gravity  are set to their original values. Now the normal load is applied on the sample using a  servo control mechanism (P-control).  After  the specified normal load is reached and the packing is at rest, the shearing phase starts by imposing a velocity on the lower ring. Variations of the shear velocity showed that shearing with 10~$\frac{\mbox{\small mm}}{\mbox{\small s}}$ yielded results, which can be considered quasi-static, i.e.,~a lower shearing rate yielded the same result. The shearing force is calculated as the sum of the forces on all walls and the bottom plate of the lower ring  in direction of the shearing. From the conducted simulations the final shear stress  is calculated as the median of the last hundred readings of the  shear stress (over a shear path of 2~mm). 
 Here, the median instead of the mean value is chosen due to its insensitivity with respect to outliers. The bulk friction coefficient is then calculated as final shear force divided by normal stress. 

\subsection{Usage of constant interparticle friction coefficients}

\begin{figure}
\centering
\parbox{0.55\textwidth}{
\includegraphics[width=0.55\textwidth]{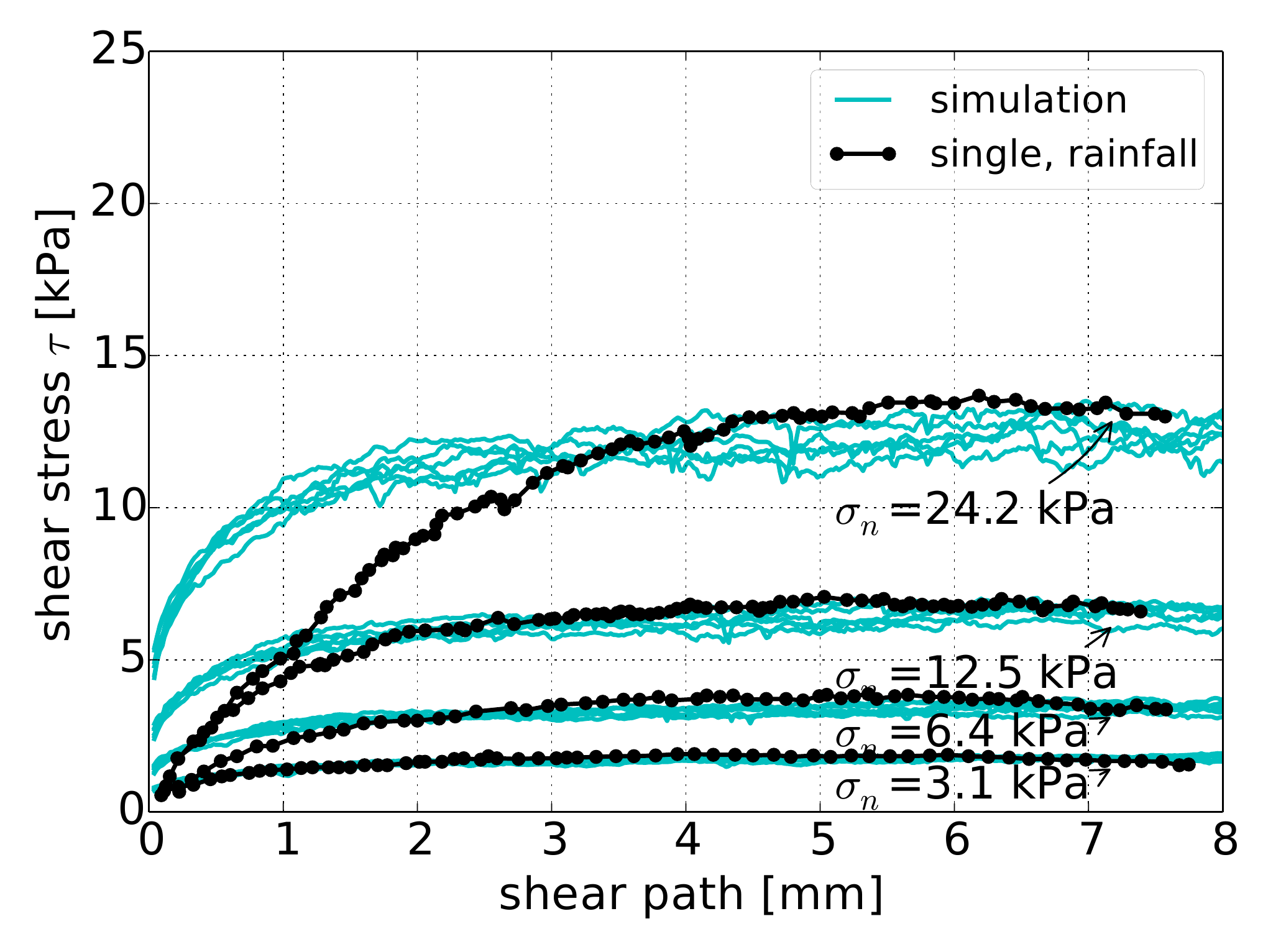}
\caption{Single spheres: Comparison of simulated shear stress ($\mu=0.2$) with experimental results of rainfall filling method (exp. results taken from \cite{Haertl2008}).}\label{fig:singleShearStress}
}
\end{figure}

In a first step, simulations with single spheres, as well as with paired spheres, will be carried out using a constant interparticle friction coefficient. Six different samples are generated to take into account the influence of small changes in initial settings on the results.
  In Figure~\ref{fig:singleShearStress}, the results for single spheres are shown using $\mu=0.2$. For all four applied normal stresses and all  six initial settings the resulting shear stress is plotted over the shear path. The experimental results for a sample generated with the rainfall filling method is obtained from \cite{Haertl2008} by plot digitalisation and thus may contain small inaccuracies. The simulations are in good accordance with the experimental results, with exception of the initial slope. 
  It is not unusual that the shearing response observed in DEM simulations is stiffer than the one seen in experiments, compare \cite{Haertl2008} and references within. 
  The numerical scatter in the results is of same order as reported by H\"artl  and Ooi, \cite{Haertl2008}, i.e.,~the coefficient of variation (CoV), which is defined as the standard deviation divided by the mean value, is below 8\%. 

\begin{figure}
\centering

\parbox{0.55\textwidth}{
\includegraphics[width=0.55\textwidth]{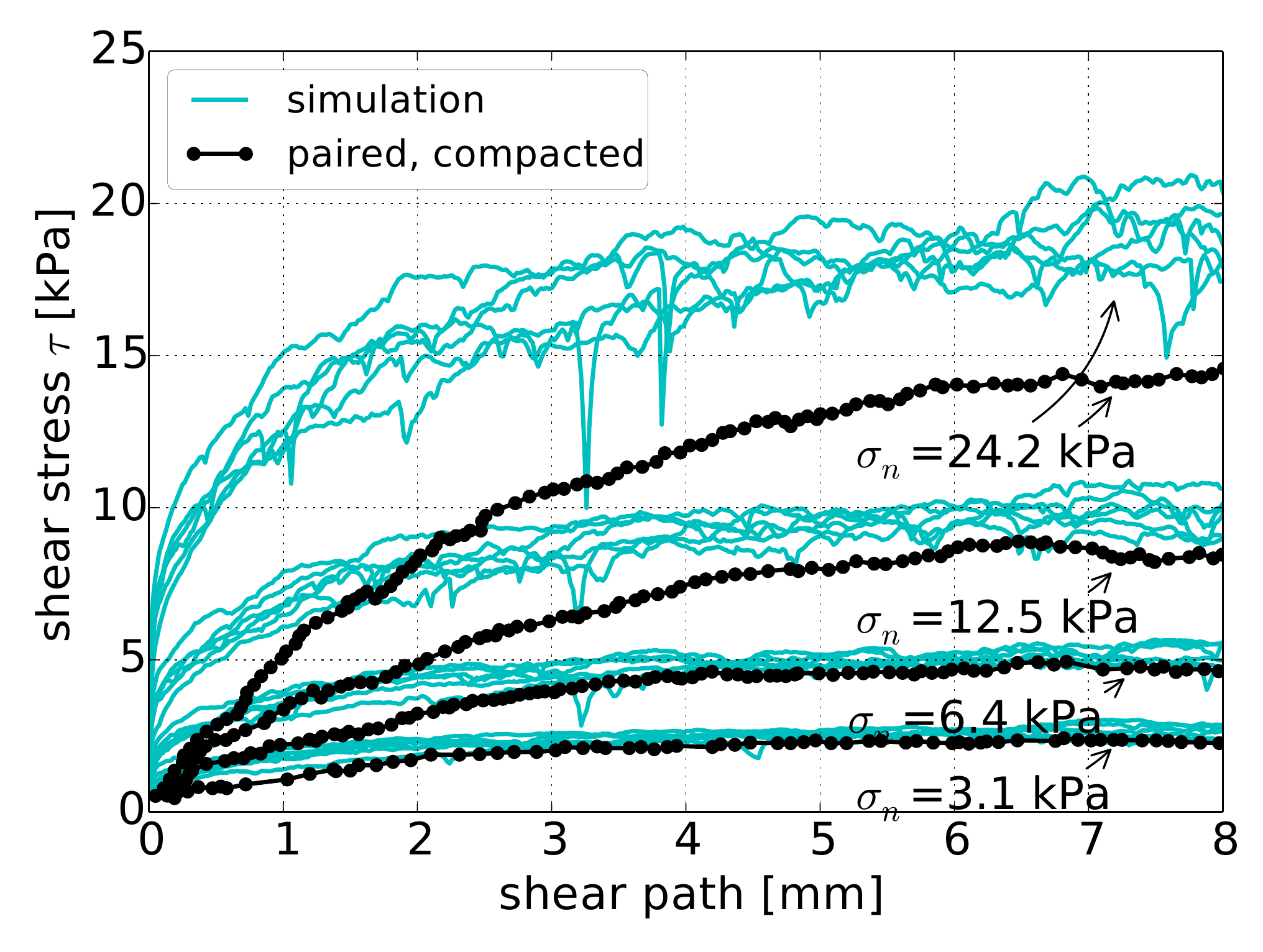}
\caption{Double spheres: Comparison of simulated shear stress ($\mu=0.2$) with experimental results of compacted filling method (exp. results taken from \cite{Haertl2011}).}\label{fig:doubleShearStress}
}
\end{figure}

The same computations are conducted for the paired spheres. 
  In Figure~\ref{fig:doubleShearStress}, the shear stress over the shear path is plotted for simulations ($\mu=0.2$) and experimental results for compacted filling. The experimental results were obtained  from \cite{Haertl2011} by plot digitalisation. Shear force plots for other filling methods, which would correspond better to the DEM sample's porosities, were not given and can thus not be used for comparison.  From the simulation data, small spikes in the plot can be seen. 
  Most of these spikes are caused by a sudden drop in the number of contacts in the sample but sometimes they are caused also by problems of the control strategy for the application of the normal stress. While for the simulation of single spheres the normal stress could be controlled with a maximal error of 1.5\%, in simulations of paired spheres deviations of up to 8\% were observed in two cases. A variation of the parameters in the control strategy showed that the influence of this problem   on the obtained bulk friction coefficient is negligible. 
   In general the simulated bulk friction coefficients are too high compared to the experimental results. 
  
\begin{figure}
\centering

\parbox{0.55\textwidth}{
\includegraphics[width=0.55\textwidth]{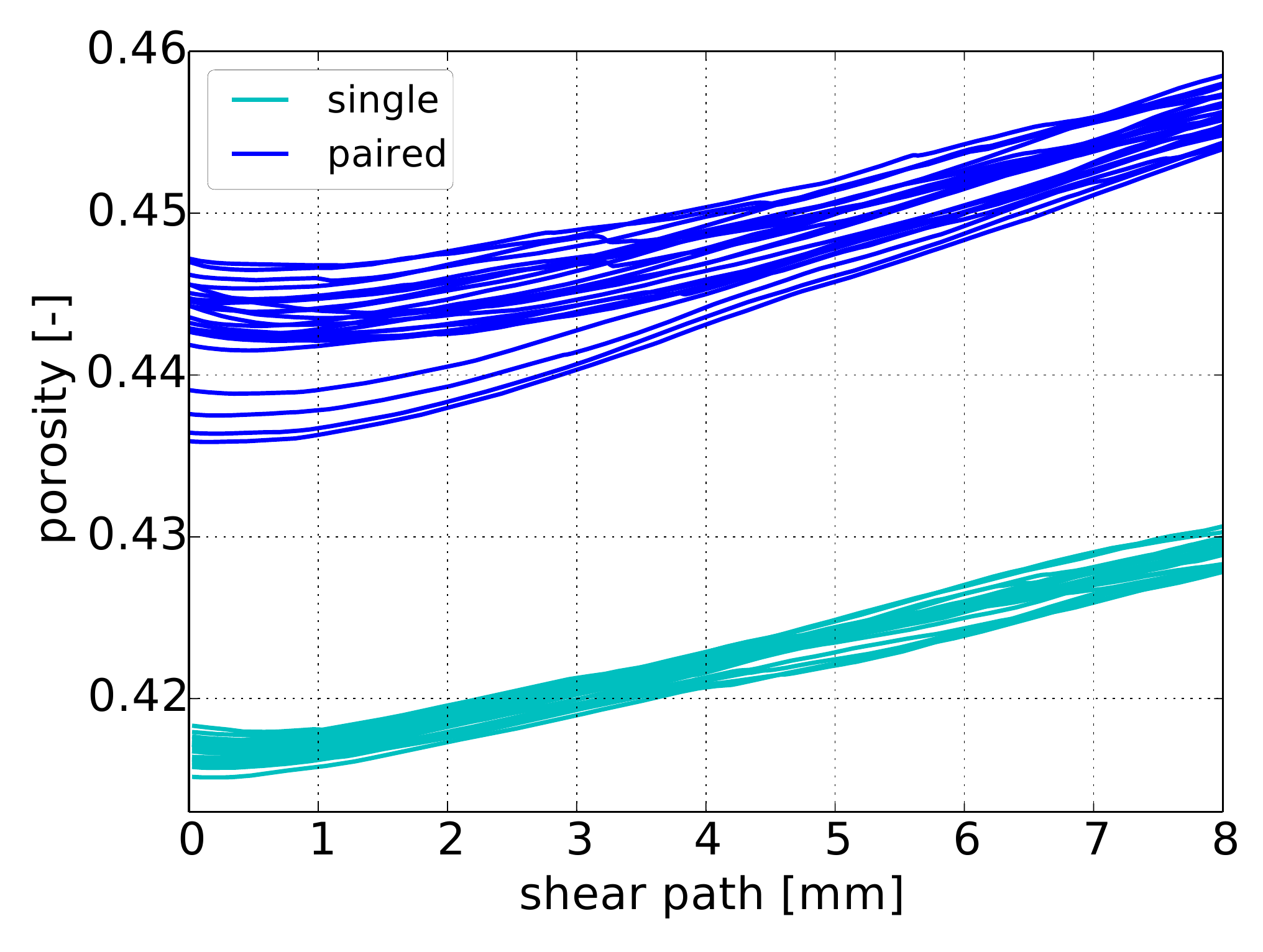}
\caption{Porosity over shear path for simulations with  ($\mu=0.2$). Comparison of single and paired spheres. }\label{fig:PoroSim}
}
\end{figure}

Figure~\ref{fig:PoroSim} shows the porosity over the shear path for simulations of both single and paired spheres.   For single spheres the initial porosities of the generated specimen varied between 0.415 and 0.419 and were thus slightly higher than the porosity of the experimental samples. The effect of looser DEM samples, compared to lab samples, was also observed by  H\"artl  and Ooi, \cite{Haertl2008}, as well as other researchers. During shearing there was compaction for the first 1~mm followed by  nearly linear dilation.  For paired spheres the initial porosities ranged from 0.436 to 0.448, thus the porosities of the generated DEM samples where in between those of the rainfall and central filling method. The compaction-dilation behaviour of the paired spheres is similar to that of single spheres. 

\begin{figure}
\centering

\parbox{0.55\textwidth}{
\includegraphics[width=0.55\textwidth]{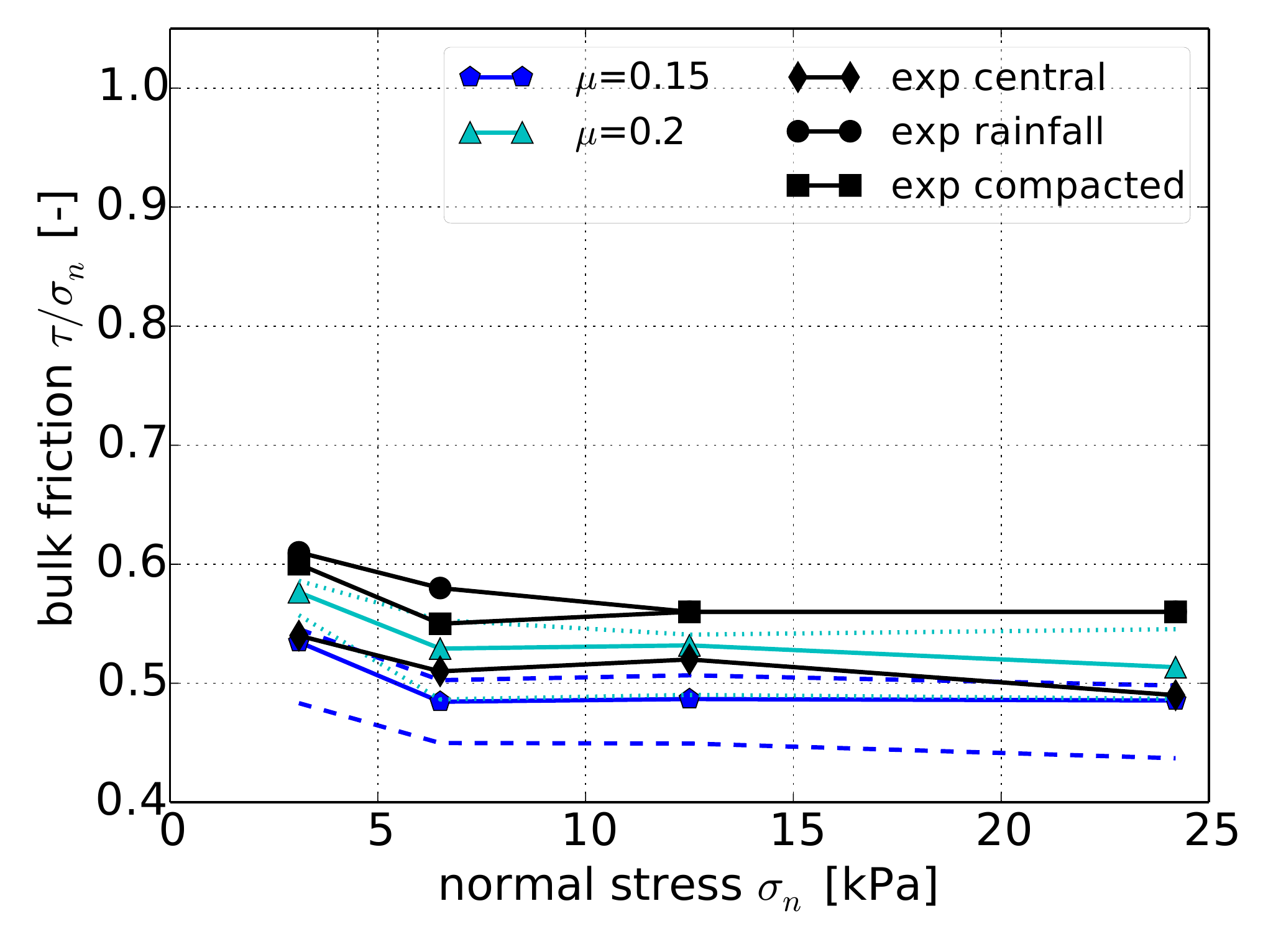}
\caption{Single spheres:  Influence of (const.) interparticle friction on bulk friction in comparison to experimental results.}\label{fig:singleBulkFrictMuConst}
}
\end{figure}
\clearpage
\begin{figure}
\centering

\parbox{0.55\textwidth}{
\includegraphics[width=0.55\textwidth]{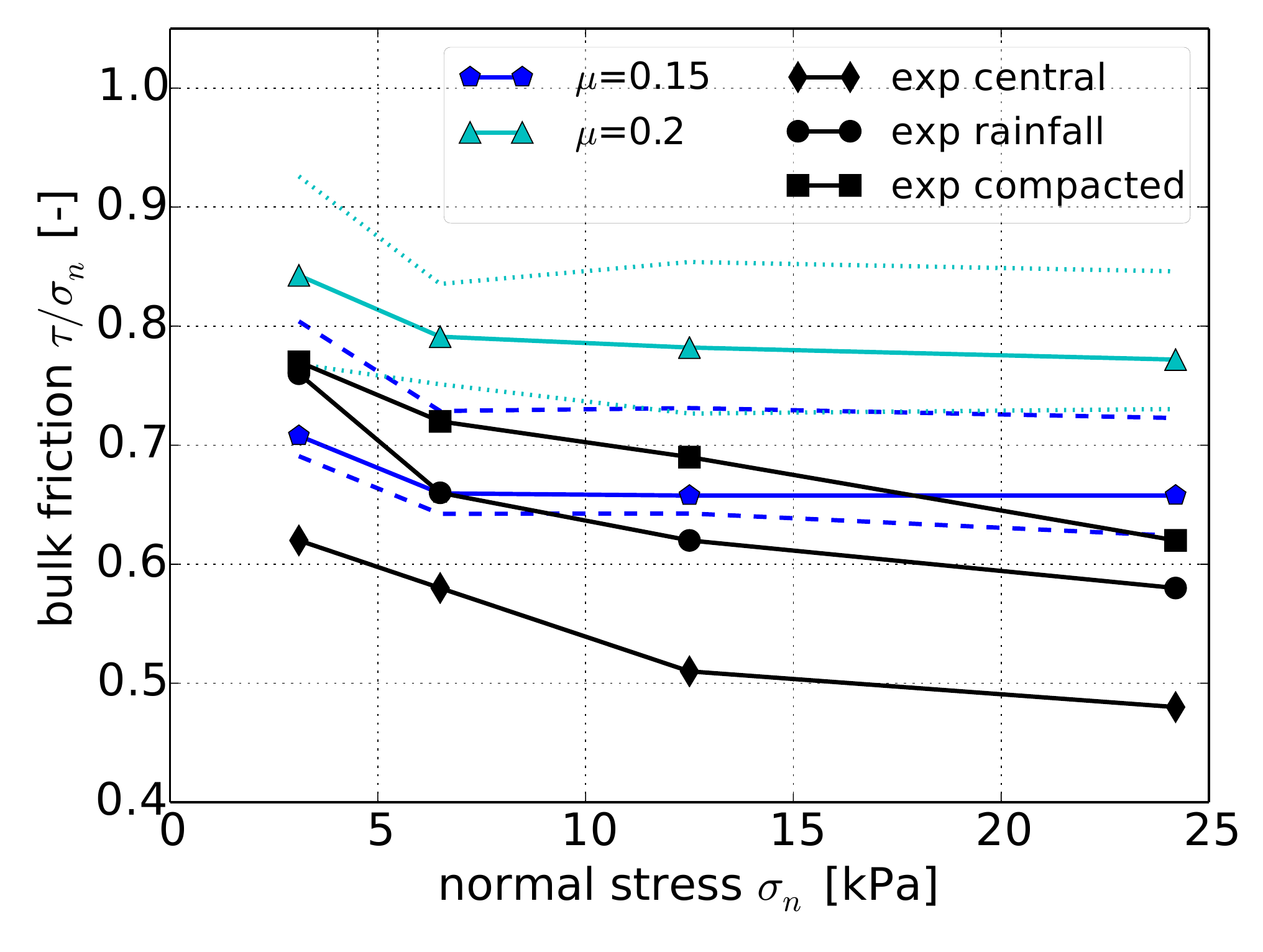}
\caption{Paired spheres:  Influence of (const.) interparticle friction on bulk friction in comparison to experimental results.}\label{fig:pairedBulkFrictMuConst}
}
\end{figure}
 In Figures~\ref{fig:singleBulkFrictMuConst} and \ref{fig:pairedBulkFrictMuConst} the  bulk friction, $\tau_f/\sigma_n$, is plotted over the applied normal stress for single and paired spheres respectively. Simulation results for $\mu=0.2$, as well as $\mu=0.15$, are shown, where the solid lines are the median of all six curves and the minimum/maximum values are plotted using dashed lines ($\mu=0.15$) or dotted lines ($\mu=0.2$). 
For the single spheres the simulation results belonging to $\mu=0.2$ lie between the experimental results of compacted/rainfall and central filling. The results for $\mu=0.15$ lie below the  experimental results. 
If the values for the lowest stress level are excluded, both  experimental and simulation results are nearly constant. As the stress dependency of the bulk friction coefficient is very small for single spheres, the usage of a constant $\mu$ yields good results.  
For paired spheres it can be noted that the simulated bulk friction obtained with $\mu=0.2$ is too high compared to the experimental results. For $\mu=0.15$ the results are between the experimental ones of compacted and rainfall filling method. Deviations to the experimental values are seen, because no dependency on the applied normal stress is present.

From these results it can be concluded that in DEM simulations with a constant interparticle friction coefficient it is not possible to obtain a stress dependency. Thus, the decay of the bulk friction with increasing normal stress, which is seen in the experimental data for paired spheres, can not be reproduced.
  H\"artl  and Ooi, \cite{Haertl2011}, also conducted DEM simulations and observed the same effect of a constant  bulk friction coefficient for both single and paired spheres. Possible explanations for them were a too simplistic contact model,  missing deviations of surface roughness or sphericity or the modelling of the tester via rigid frictional bounds. 
  The authors of this study assume, that the observed normal stress dependency of the bulk friction coefficient is caused by  tribological effects. Thus, a more tribological tangential contact law is implemented in DEM.

\subsection{Stress dependent  interparticle friction -- a parameter study}

To obtain a stress dependency in the bulk friction of the granular material, the contact model with pressure dependent interparticle friction will be used, as defined in \eqref{eq:pdfPopov}. As already mentioned no measurement data for glass-glass contact is available for the calibration of the model.  Therefore,   in this subsection the influence of the three model parameters, $\mu_0, c_1, c_2$ , on the calculated bulk friction will be investigated. 

\begin{figure}
\centering
\parbox{0.55\textwidth}{
\includegraphics[width=0.55\textwidth]{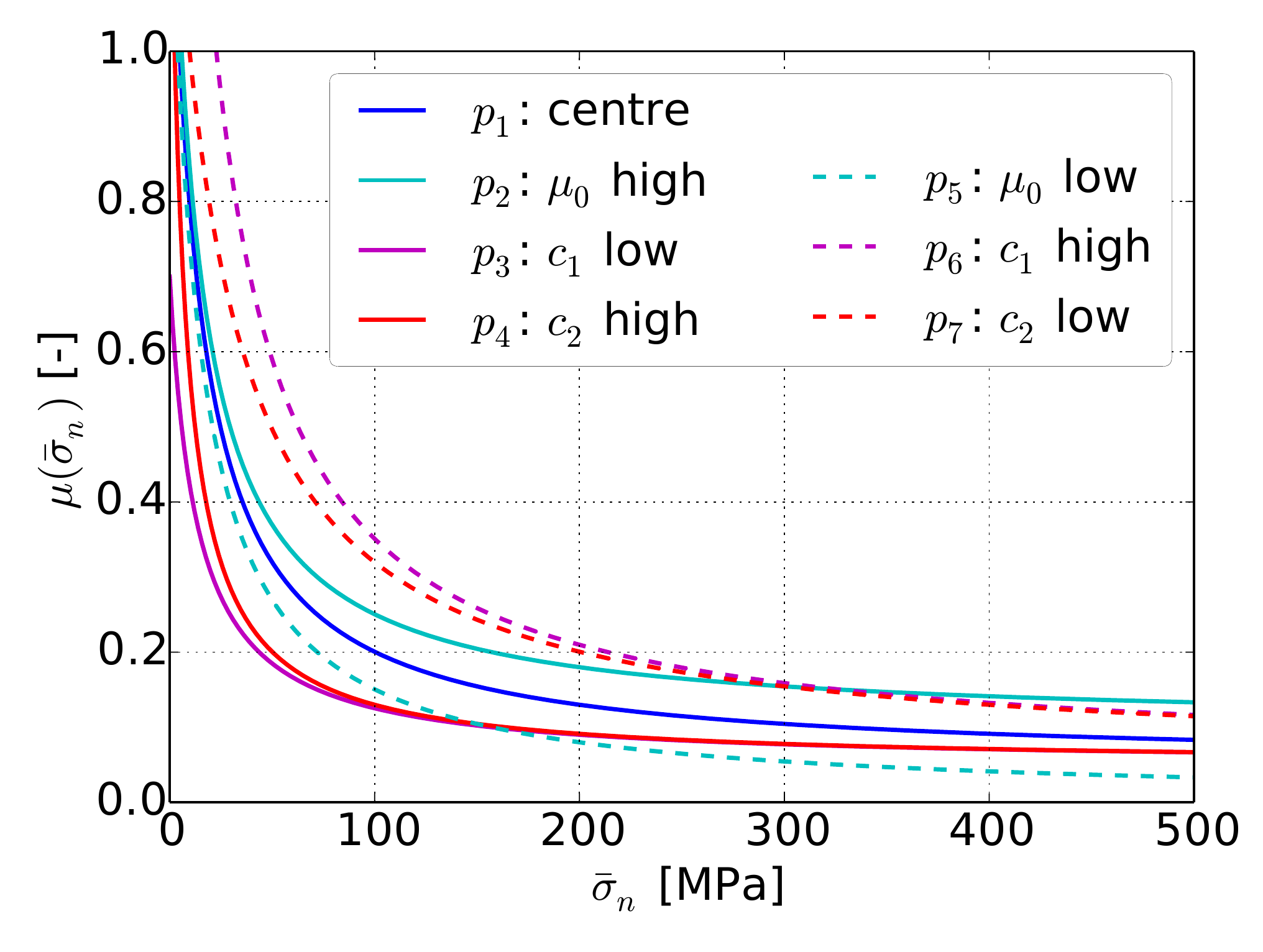}
\caption{Influence of parameters within stress dependent friction model \eqref{eq:pdfPopov}.}\label{fig:PopovParam}
}
\end{figure}

Regarding the influence of the model parameters on the interparticle friction, it is worth noting that the function is bounded: 
\begin{subequations}
\begin{align}\label{eq:modelBounds}
&\mbox{upper bound: } \mu(0)= \mu_0 + c_1 \\
&\mbox{lower bound: }\lim_{\bar{\sigma}_n \rightarrow \infty}  \mu(\bar{\sigma}_n)= \mu_0
\end{align}
\end{subequations}
In the following, one point in the parameter space is chosen as central point, $p_1$,  and for each of the three parameters one lower and one higher value will be considered. The resulting seven sets of parameters, $p_1$, \ldots, $p_7$, are listed in Table~\ref{tab:PopovPar}, and the corresponding graphs of the model are plotted in Figure~\ref{fig:PopovParam}.
  Obviously, a variation of the parameter $\mu_0$ causes a vertical shift of the graph,  compare the curves of $p_1$, $p_2$ and $p_5$ in Figure~\ref{fig:PopovParam}. For higher stresses the graphs become nearly linear. The parameters $c_1$ and $c_2$ influence the slope of these lines, as well as range of stresses, where linear behaviour is a good approximation of the graphs. %

\begin{table}
\begin{tabular}{llllllll}
& $p_1$&$p_2$&$p_3$&$p_4$&$p_5$&$p_6$&$p_7$\\
\hline
$\mu_0$ [-]& 0.005&0.1& 0.05&0.05&0.0&0.05&0.05\\
$c_1$ [-]&1.3&1.3&0.65&1.3&1.3&2.6&1.3\\
$c_2$ [-]&7.64e-8&7.64e-8&7.64e-8&1.53e-7&7.64e-8&7.64e-8&3.82e-8\\
\hline
\end{tabular}
\caption{Parameter sets to be compared for the stress dependent friction model.}\label{tab:PopovPar}
\end{table}

 \begin{figure}
\centering
\parbox{0.55\textwidth}{
\includegraphics[width=0.55\textwidth]{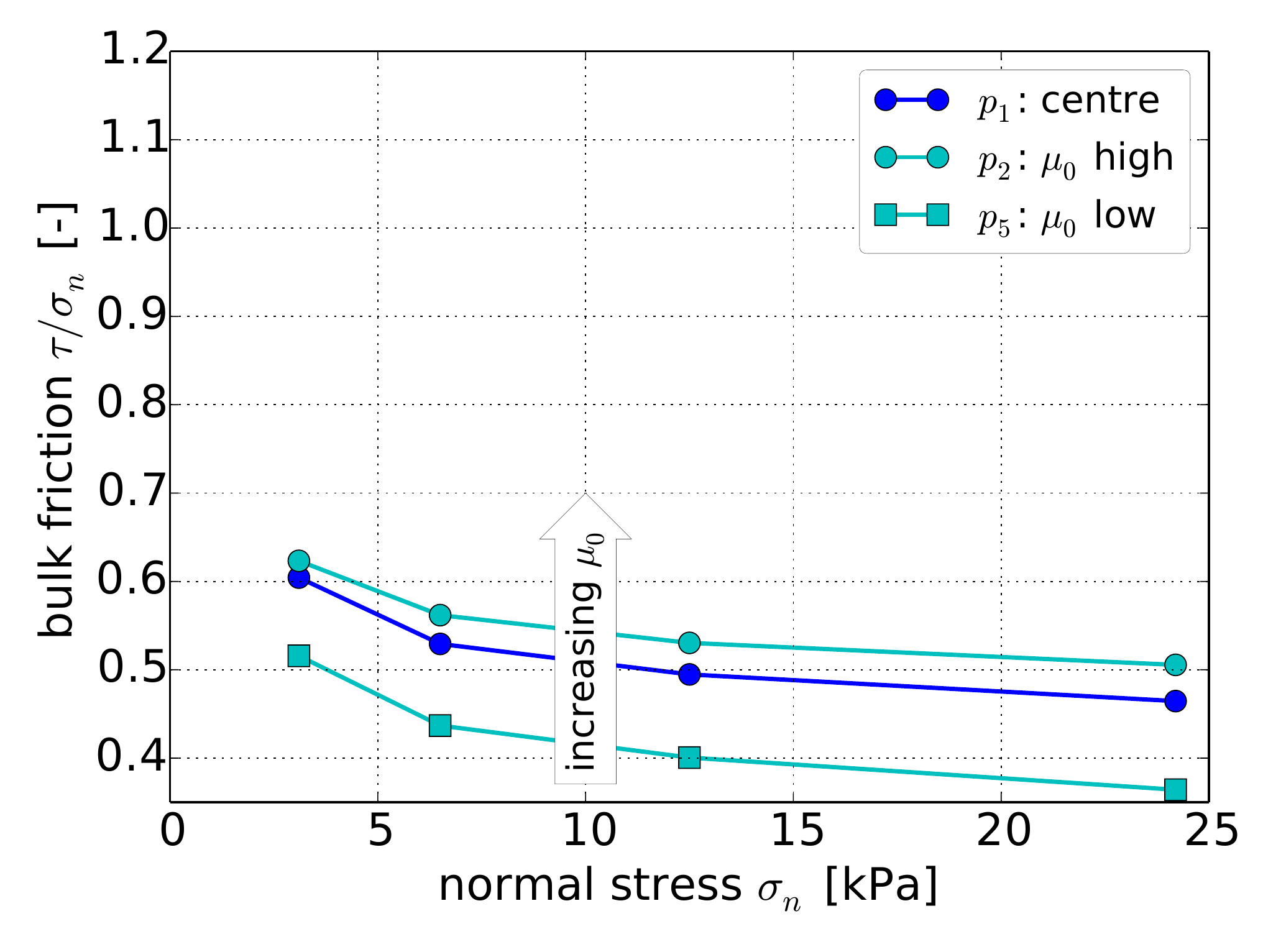}
\caption{Single spheres:  Influence of parameter $\mu_0$ in stress dependent interparticle friction on bulk friction.}\label{fig:singleBulkFrictPDFVar_mu0}
}
\end{figure}
\clearpage
\begin{figure}

\parbox{0.55\textwidth}{
\includegraphics[width=0.55\textwidth]{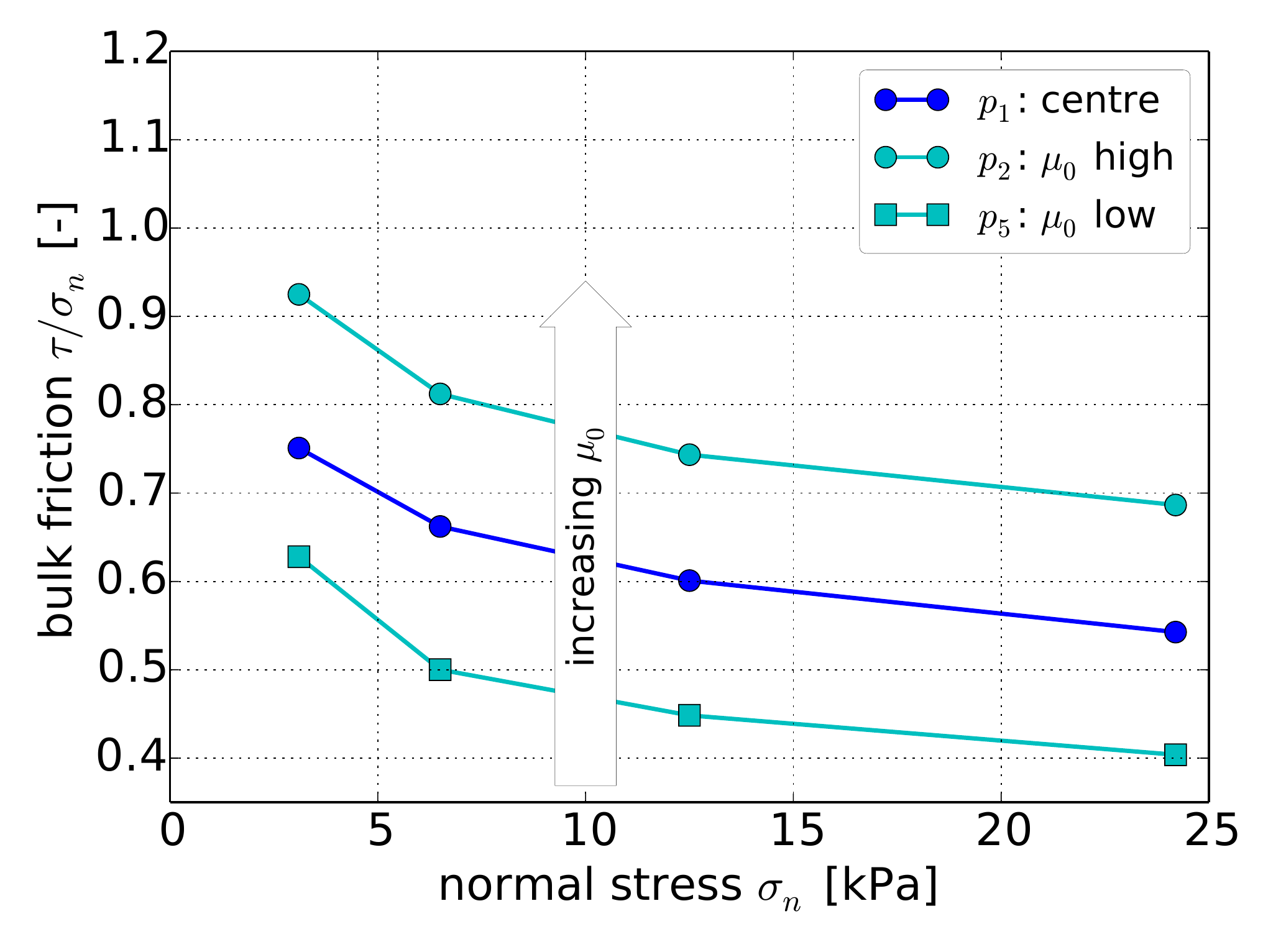}
\caption{Paired spheres:  Influence of parameter $\mu_0$ in stress dependent interparticle friction on bulk friction.}\label{fig:pairedBulkFrictPDFVar_mu0}
}
\end{figure}

\begin{figure}
\centering
\parbox{0.55\textwidth}{
\includegraphics[width=0.55\textwidth]{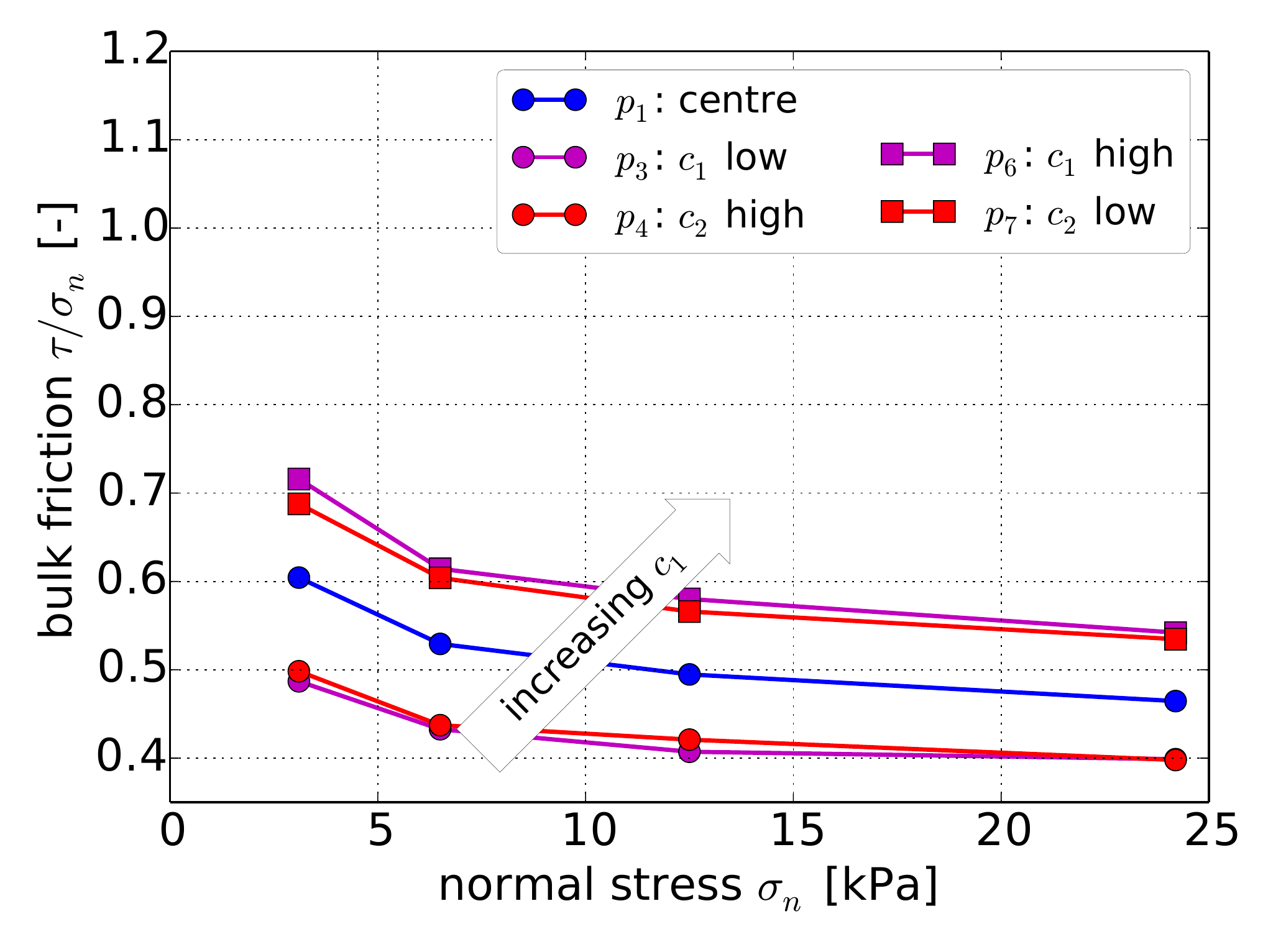}
\caption{Single spheres:  Influence of parameters $c_1, c_2$ in stress dependent interparticle friction on bulk friction.}\label{fig:singleBulkFrictPDFVar_c1c2}
}
\end{figure}

\begin{figure}
\centering

\parbox{0.55\textwidth}{
\includegraphics[width=0.55\textwidth]{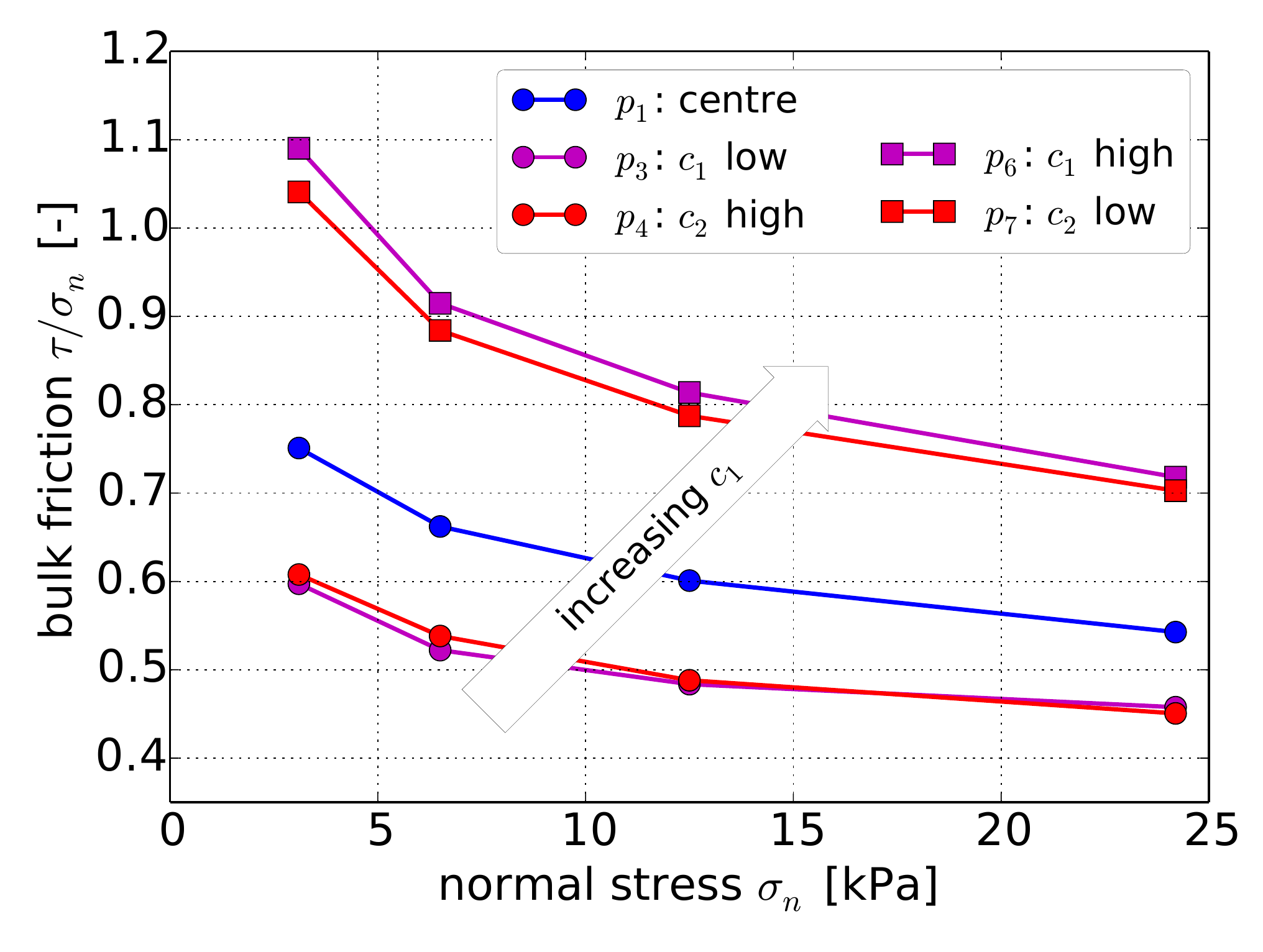}
\caption{Paired spheres:  Influence of parameters $c_1, c_2$ in stress dependent interparticle friction on bulk friction.}\label{fig:pairedBulkFrictPDFVar_c1c2}
}
\end{figure}

Simulations with the seven sets of parameters were conducted for single spheres as well as for paired spheres. At first, a  change of parameter $\mu_0$ is considered ($p_1, p_2, p_5$). The corresponding results are presented in Figures~\ref{fig:singleBulkFrictPDFVar_mu0} and \ref{fig:pairedBulkFrictPDFVar_mu0}, where the bulk friction over the normal stress is plotted. Only the median of all six conducted simulations is shown.  
 It can be seen that with increasing $\mu_0$ the bulk friction coefficient curves shifts upwards, while nearly no changes in the slope of the curve occur. Both single and paired spheres show the same effect. However, its extent is much bigger for paired spheres than for single spheres. 
 Next, changes in parameters $c_1$ ($p_3, p_6$) and $c_2$ ($p_4, p_7$) are investigated. The results of the simulations are plotted in Figures~\ref{fig:singleBulkFrictPDFVar_c1c2} and \ref{fig:pairedBulkFrictPDFVar_c1c2}. When parameter $c_1$ increases, the bulk friction coefficient curve shifts upwards, and a growth of its slope can be observed. Again, these effects are equal for single and paired spheres, while they are more apparent with paired spheres. A variation of parameter $c_2$ shows the inverse effect, thus the described behaviour can be seen for decreasing $c_2$. 
   When comparing $p_3$ with $p_4$ or $p_6$ with $p_7$, nearly no differences in the bulk friction coefficient curves can be seen. The corresponding graphs of the stress dependent internal friction were almost identical for stresses above 200~MPa, see Figure~\ref{fig:PopovParam}. Thus, it can be concluded that contacts with small stresses barely contribute to the overall behaviour of the granular material.  
  Of course in all these considerations only the median of six simulations were used, so that the behaviour of single simulation runs may deviate.
  
 In general, it can be said that the effect that the parameters have on the interparticle friction coefficient curve is similar to their effect on the resulting bulk friction coefficient curve. 
 In the considered case it is sufficient to keep either $c_1$ or $c_2$ as a parameter in the model, while the other can be set constant. From now on,  only $\mu_0$ and $c_1$ are considered further for a calibration. 
\subsection{Stress dependent  interparticle friction -- parameter fitting}

The knowledge of the influence of both model parameters makes it now possible to calibrate the model and compare its results to the experimental data. Of course, there exist different methods which could be used for the calibration, e.g.,~numerical optimisation or linear regression models used on basis of Design or Experiments (DoE). 

In this work, a simpler approach is chosen. Simulations conducted with the central point in the parameter study, $p_1$,   already  roughly showed the wanted behaviour. This parameter set was initially obtained by a small number of `try and error' simulation runs.   Starting from this good initial guess for the parameter set, a calibration of parameter $c_1$ will be used to achieve a  slope similar to the experimental data. Then parameter $\mu_0$ will be adapted to shift the curve vertically in the right position. 

For this approach, the data for paired spheres will be used, and the experimental data for the rainfall filling is considered. For the calibration no additional simulation runs are conducted, but the data from the parameter study will be used. When the values for the lowest normal stress level are excluded, the median of the resulting bulk friction coefficient curves is roughly linear. For each data set a line is fitted though the bulk friction values using the least squares approach. In this way, the data on the slope of the lines and their intercepts are obtained. 
  At first, parameter $c_1$ is estimated using parameter sets $p_1, p_3$ and $p_6$. A linear interpolation between the slopes of the fitted lines is used to match the slope of the experimental data. Now, when $c_1$ is fixed,the resulting change in height is estimated, using another linear interpolation between the intercepts of those lines. %
 An analog procedure is used for parameter $\mu_0$ (using parameter sets $p_1, p_2$ and $p_5$) to shift the curve vertically in the right position. The obtained parameter set is $\mu_0=0.081$ and $c_1=0.841$ while $c_2=7.63e-8$ was considered constant. 
 For an arbitrary starting point of the calibration, the described approach is surely too simple. In the current situation with a given  good starting point for the parameters, the results turned out to be satisfying.

\subsection{Stress dependent  interparticle friction -- comparison with experiments}

\begin{figure}
\centering

\parbox{0.55\textwidth}{
\includegraphics[width=0.55\textwidth]{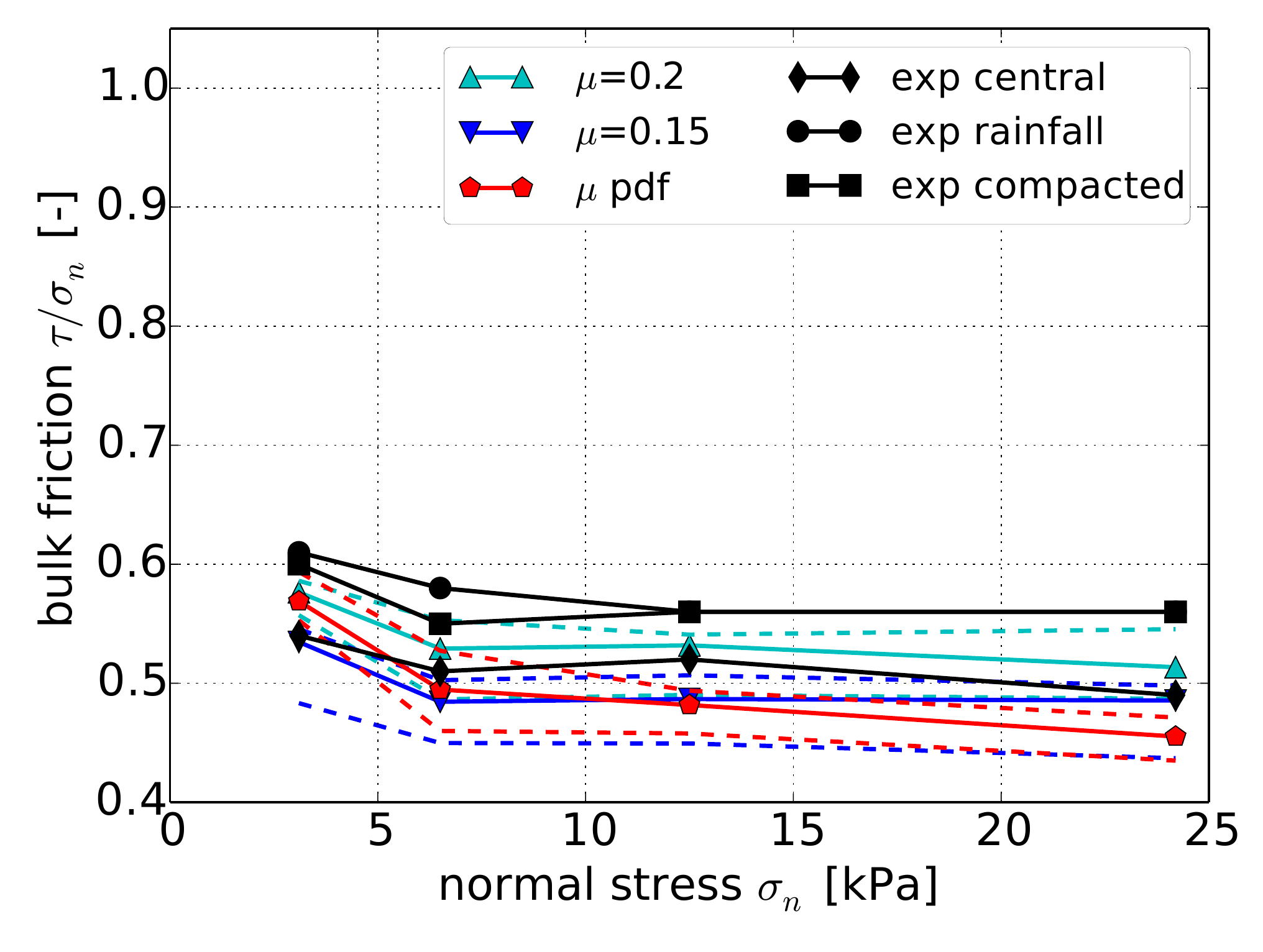}
\caption{Single spheres:  Comparison of simulation results for constant  and pressure dependent interparticle friction ($\mu$ pdf) with experiments.}\label{fig:singleBulkFrictPDF-fittedvsMu02}
}
\end{figure}

\begin{figure}
\centering

\parbox{1.05\textwidth}{
\includegraphics[width=0.5\textwidth]{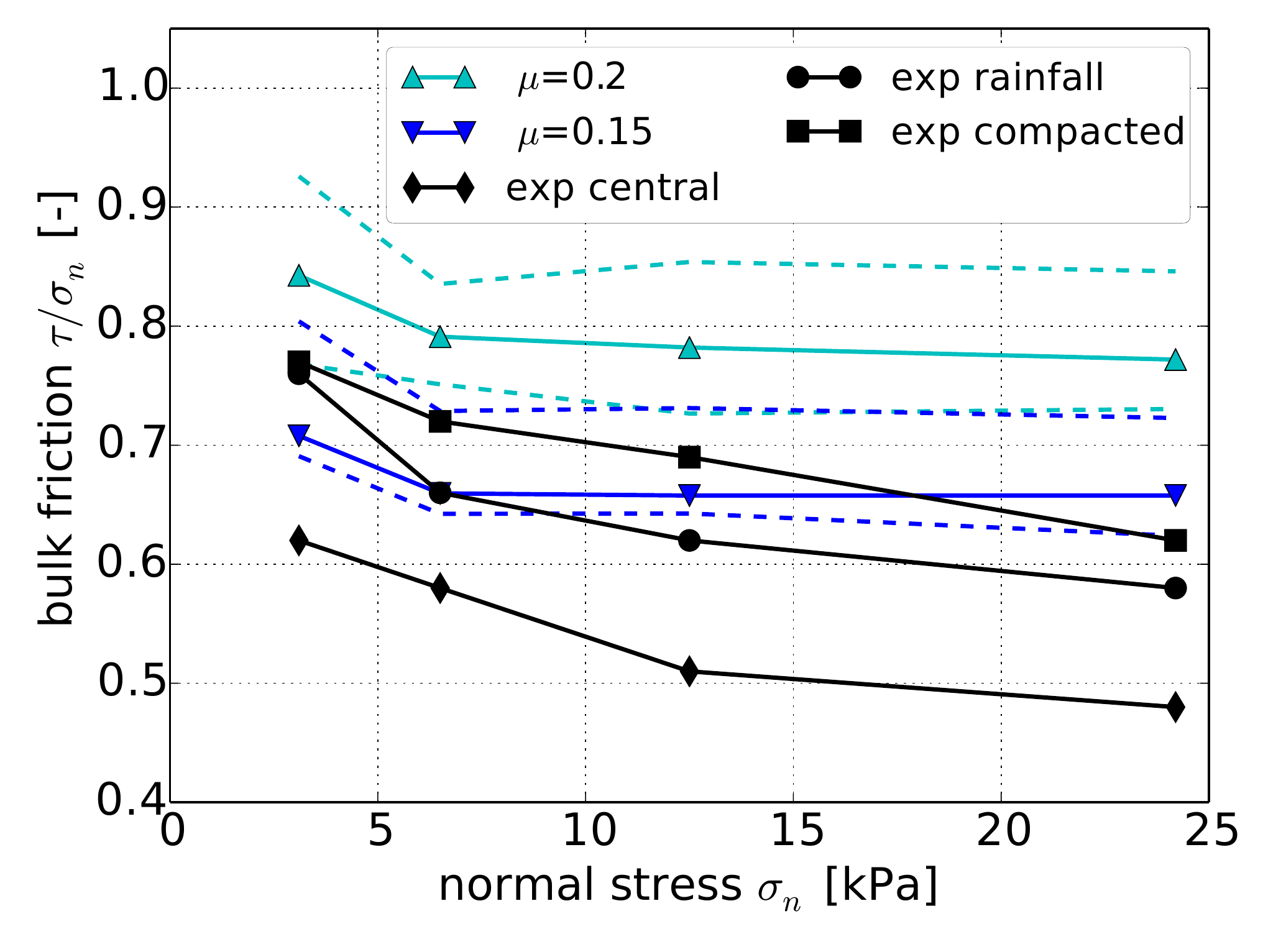}
\includegraphics[width=0.5\textwidth]{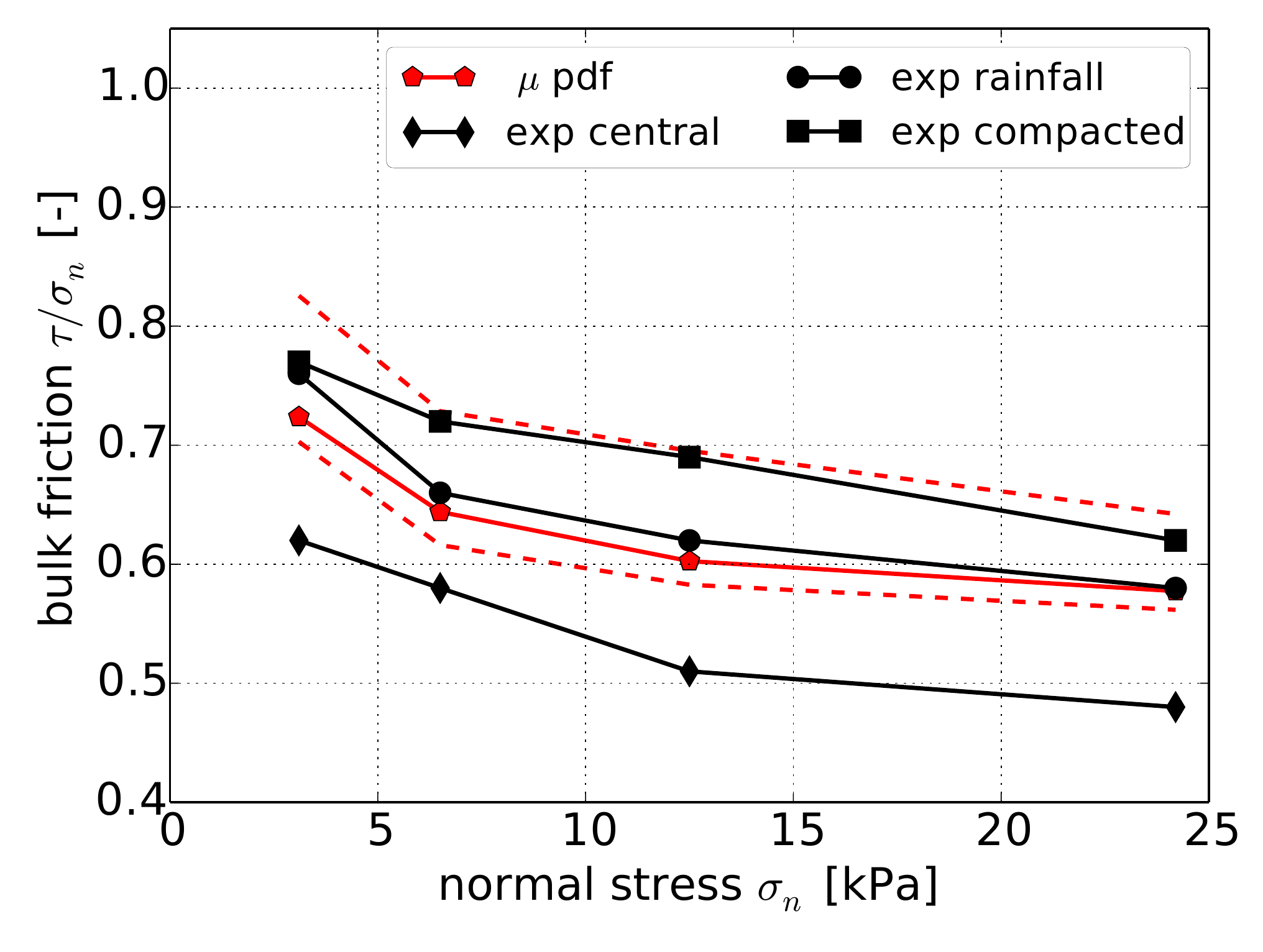}
\caption{Paired spheres:  Comparison of simulation results for constant (left) and pressure dependent interparticle friction, $\mu$ pdf, (right) with experiments.}\label{fig:pairedBulkFrictPDF-fittedvsMu02}
}\\

\end{figure}

Simulations for single and paired spheres are conducted using the obtained parameter set. In Figures~\ref{fig:singleBulkFrictPDF-fittedvsMu02}, and \ref{fig:pairedBulkFrictPDF-fittedvsMu02} the obtained bulk friction coefficient over the normal stress is plotted for the simulations with  constant interparticle friction ($\mu=0.15, \mu=0.2$), pressure dependent interparticle friction ($\mu$ = pdf) and experimental results. Again, the median of all six curves and the minimum/maximum values are plotted using dashed lines. For single spheres the best approximation of the experimental data is obtained with $\mu=0.2$, while the  results for $\mu=0.15$ and  stress dependent simulations are slightly lower than the experiments. The stress dependency of the bulk friction coefficient is negligible both for experiments and  simulations. 
As described before, for paired spheres the simulations with constant interparticle friction give a bad approximation of experimental data. 
Of the two approximations, the simulations with $\mu=0.15$ agree better with the experimental data. 
 Therefore,  the results with $\mu=0.2$ will not be considered further. The simulations with stress dependent interparticle friction agree well with the experiments. The slope of the median curve fits to the rainfall data, which was used for model calibration. Thus, the stress dependency of the  bulk friction seen in the experiments can be reproduced using a stress dependent interparticle friction. 

\begin{figure}
\centering 
\parbox{0.55\textwidth}{

\includegraphics[width=0.55\textwidth]{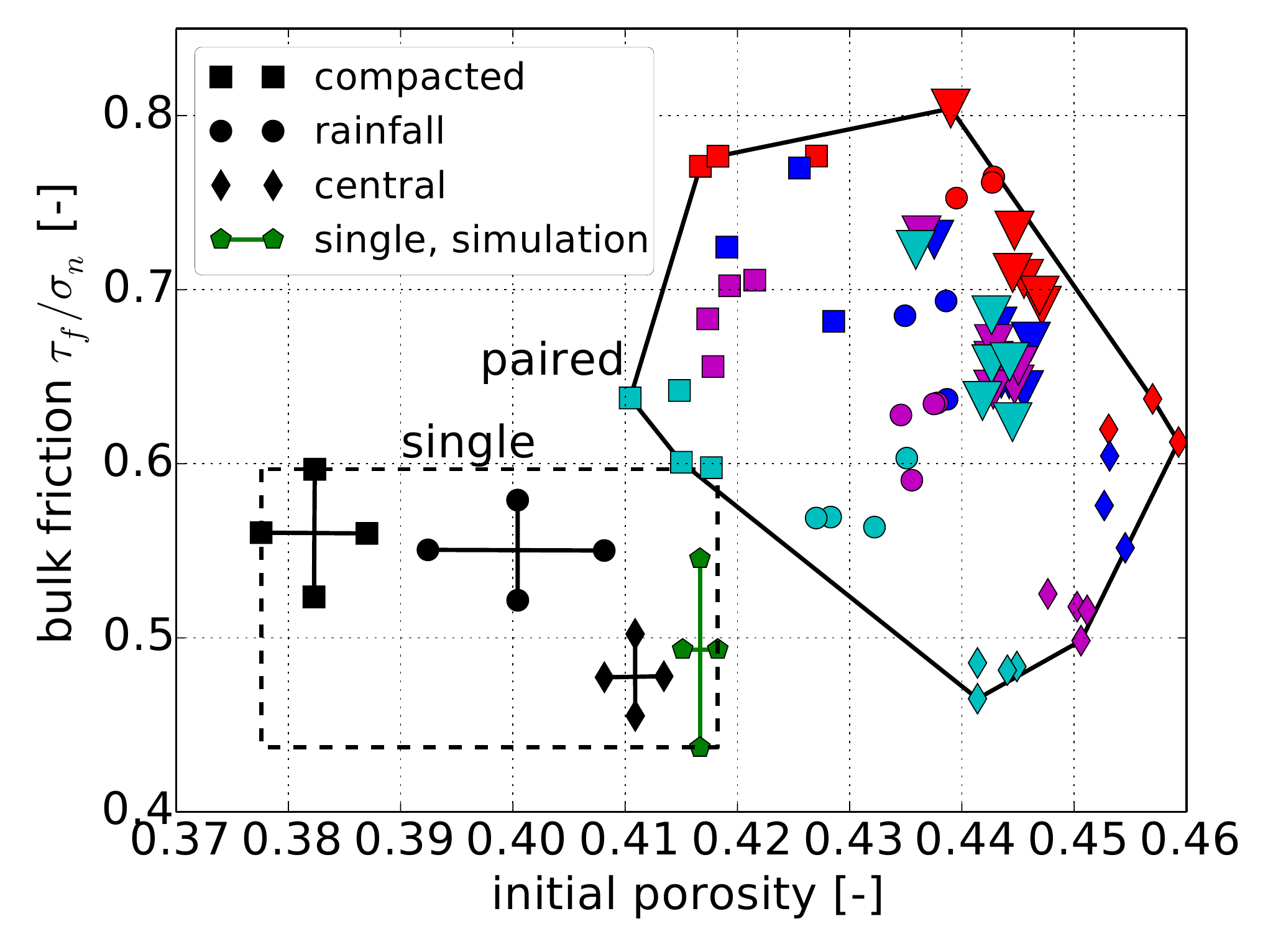}
\caption{Comparison of experimental and simulation results using $\mu=0.15$.  Marker types corresponding to filling of paired spheres; compacted: square, rainfall: circle, central: diamond, simulation paired: triangle. Colouring of applied normal stresses: 3.1~kPa: red, 6.4~kPa: blue, 12.5~kPa: magenta, 24.2~kPa: cyan.}\label{fig:PoroExpSimMu015}
}
\end{figure}

\begin{figure}
\centering

\parbox{0.55\textwidth}{

\includegraphics[width=0.55\textwidth]{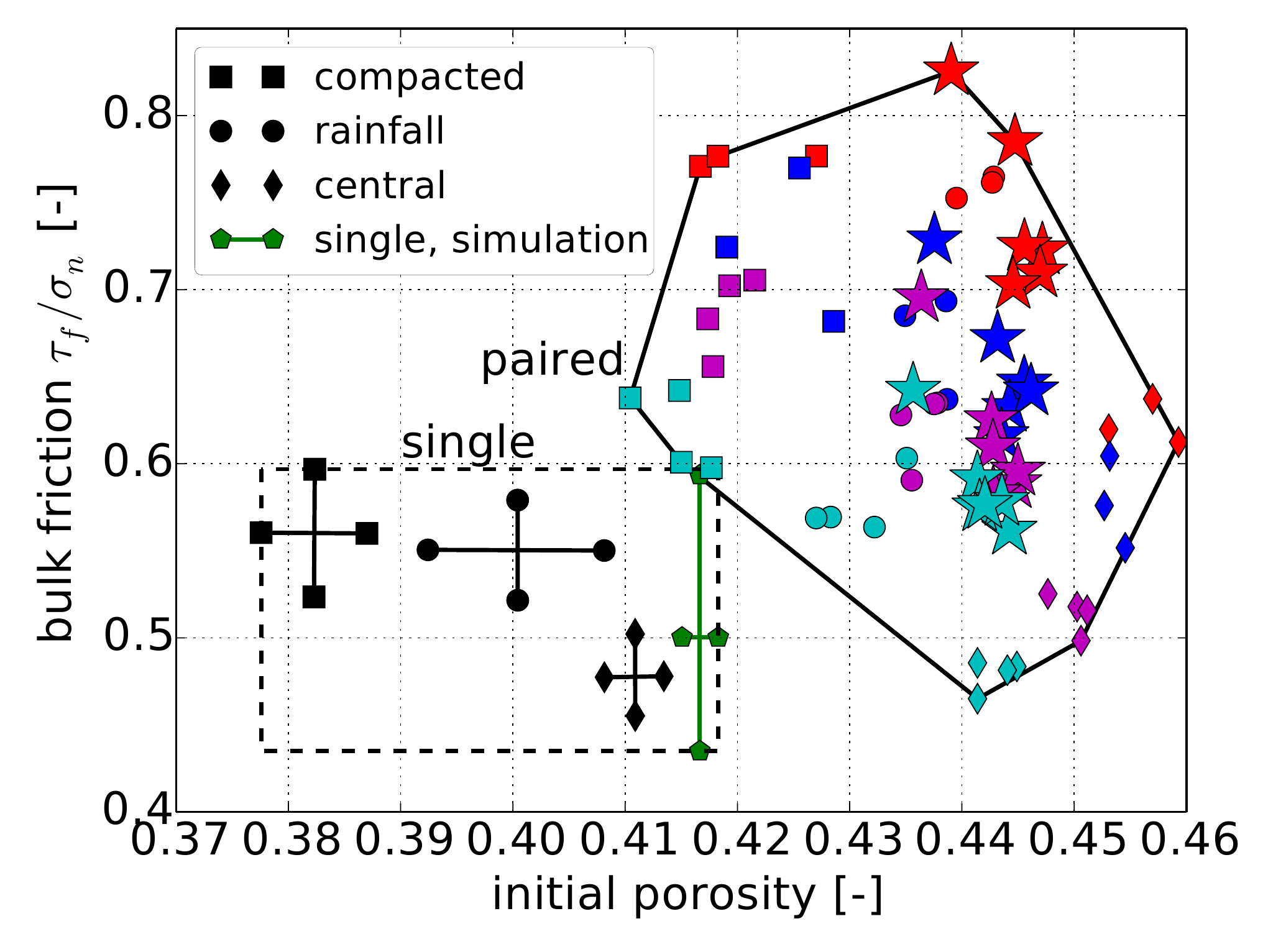}
\caption{Comparison of experimental and simulation results using stress dependent interparticle friction.  Marker types corresponding to filling of paired spheres; compacted: square, rainfall: circle, central: diamond, simulation paired: star. Colouring of applied normal stresses: 3.1~kPa: red, 6.4~kPa: blue, 12.5~kPa: magenta, 24.2~kPa: cyan.}\label{fig:PoroExpSimPdf}
}

\end{figure}

For a second comparison the simulation results are plotted together with the experimental results as bulk friction over initial porosity, in Figure~\ref{fig:PoroExpSimMu015} for $\mu=0.15$ and in Figure~\ref{fig:PoroExpSimPdf}  for  stress dependent interparticle friction. 
For single spheres the simulation results for $\mu=0.15$ and pressure dependent $\mu=$  are not too different. Only the bulk friction values for the lowest level of normal stress are higher with stress dependent interparticle friction. Considering paired spheres, in Figure~\ref{fig:PoroExpSimMu015} it can be seen clearly that the simulations with $\mu=0.15$ show no stress dependency in the results. The obtained bulk friction values for $\sigma_n=6.4, 12.2, 24.4$~kPa are scattered between 0.6 and 0.7 in no order.  On the contrary, in Figure~\ref{fig:PoroExpSimPdf}, the usage of stress dependent interparticle friction leads to a decay of bulk friction with increasing normal stress. Although some scatter is present, in  experimental  as well as in simulation results,  the prediction quality of the simulation is considerably improved compared to the usage of constant $\mu$.  

\subsection{Stress dependent  interparticle friction -- further model analysis}

\begin{figure}
\centering

\parbox{\textwidth}{
\includegraphics[width=\textwidth]{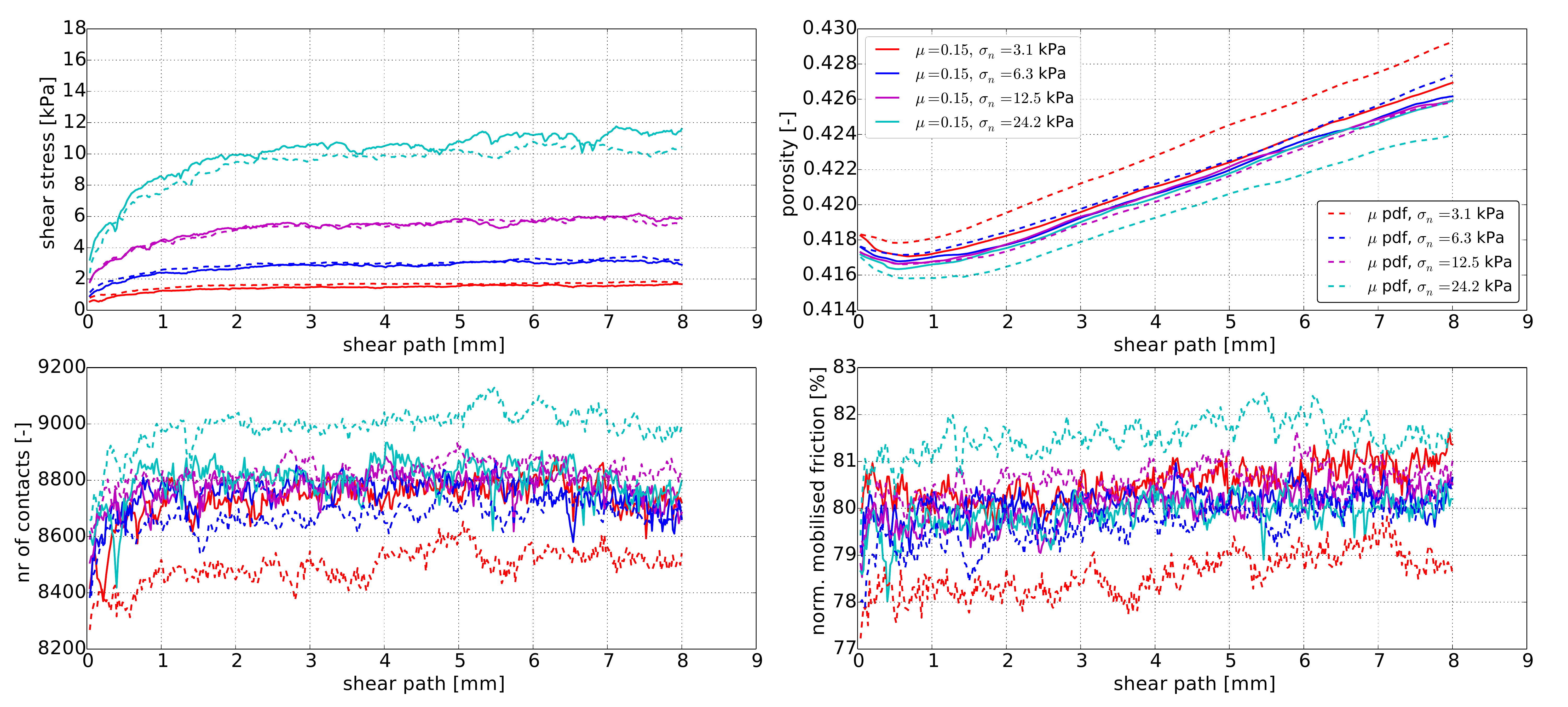}
\caption{Single spheres:  shear stress, porosity, number of contacts and normalised mobilised friction over shear path for simulations with $\mu=0.15$ and $\mu$ pdf. }\label{fig:single2x2}
}
\end{figure}

\begin{figure}
\centering

\parbox{\textwidth}{
\includegraphics[width=\textwidth]{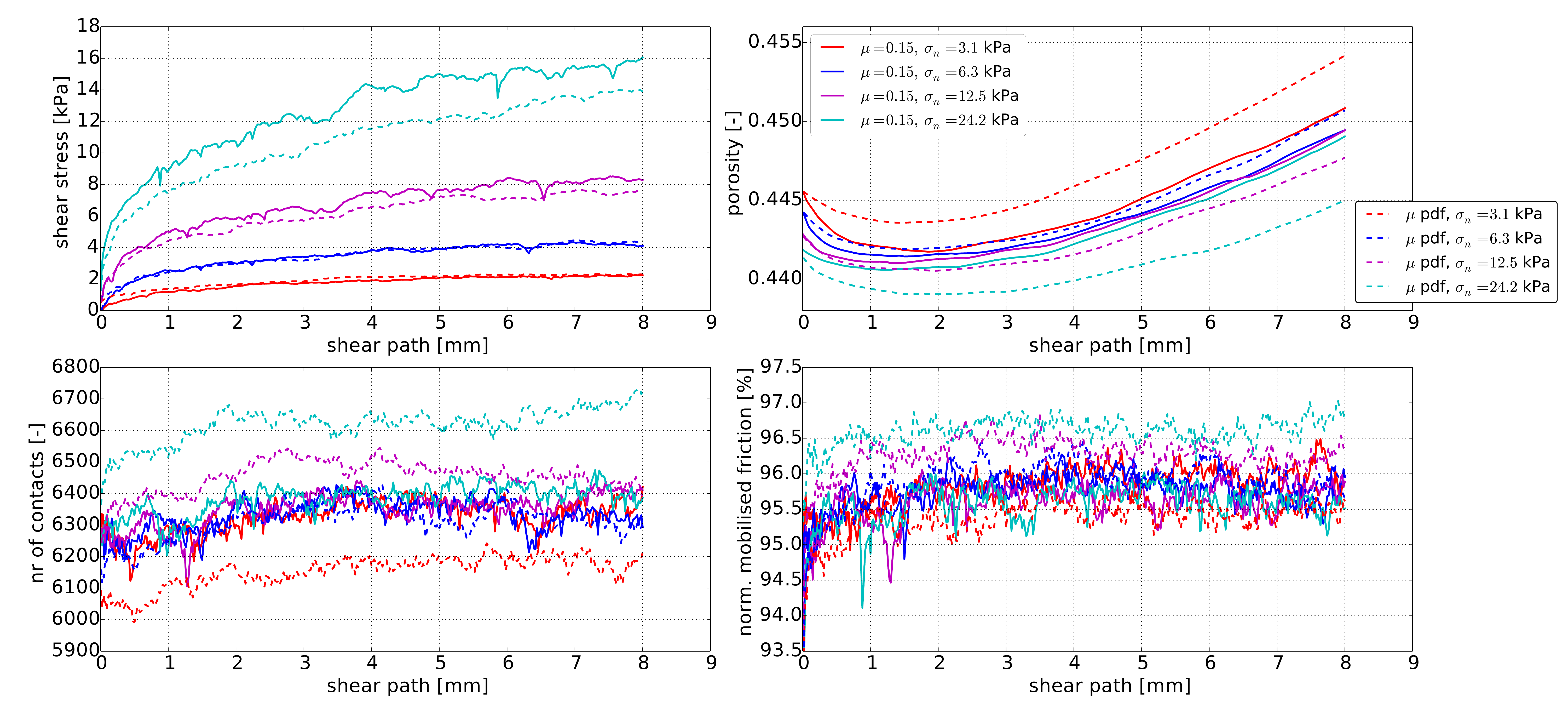}
\caption{Paired spheres:   shear stress, porosity, number of contacts and normalised mobilised friction over shear path for simulations with $\mu=0.15$ and $\mu$ pdf.}\label{fig:double2x2}
}
\end{figure}

To gain a deeper understanding of the effect of the proposed model on the simulated material behaviour, detailed plots are provided.    The shear stress, porosity, number of contacts and normalised mobilised friction over the shear path for both $\mu=0.15$ and pressure dependent $\mu$ are shown for single (Figure~\ref{fig:single2x2}) and paired spheres (Figure~\ref{fig:double2x2}). 
The normalised mobilized friction is defined as
\vskip-.6cm
\begin{eqnarray}\label{eq:MobFrict}
  m_f=\frac{1}{N_c}\sum_{k=1}^{N_c} \frac{F^k_t}{F^k_n \mu}\; .
\end{eqnarray}
 Normalisation takes place via the devision by $\mu$. Thus, its values lie between 0 and 1 (instead of 0 and $\mu$). This makes it  applicable also for the model with stress dependent $\mu$. 
 In both Figures the  shear stress versus shear path is plotted for single and for paired spheres. %
 Simulation (as well as experimental) results do not show the classical shear path - shear stress relation of dense packings, including a peak shear stress followed by a reduction before the final value is reached. Nevertheless, for simulations with constant $\mu$ the obtained packings must be rather dense, as the effect of normal load on porosity is rather small. Also, the level of applied normal load has only  little influence on the  number of contacts and the mobilized friction. 
  The behaviour seen from simulations with pressure dependent $\mu$ is different, though. The different levels of applied normal load directly change the distribution of interparticle friction values. Therefore, a stronger influence of the applied normal load on the initial porosity as well as on the dilation  behaviour is observed. The change in initial porosity of the sample is naturally accompanied with a change in the overall number of contacts and also the mobilised friction changes slightly. 
 From both Figures~\ref{fig:single2x2} and \ref{fig:double2x2}, it can be seen that the pressure dependency of $\mu$ causes severe changes in the key properties of the packing.
 
\begin{figure}
\centering

\parbox{0.48\textwidth}{
\includegraphics[width=0.48\textwidth]{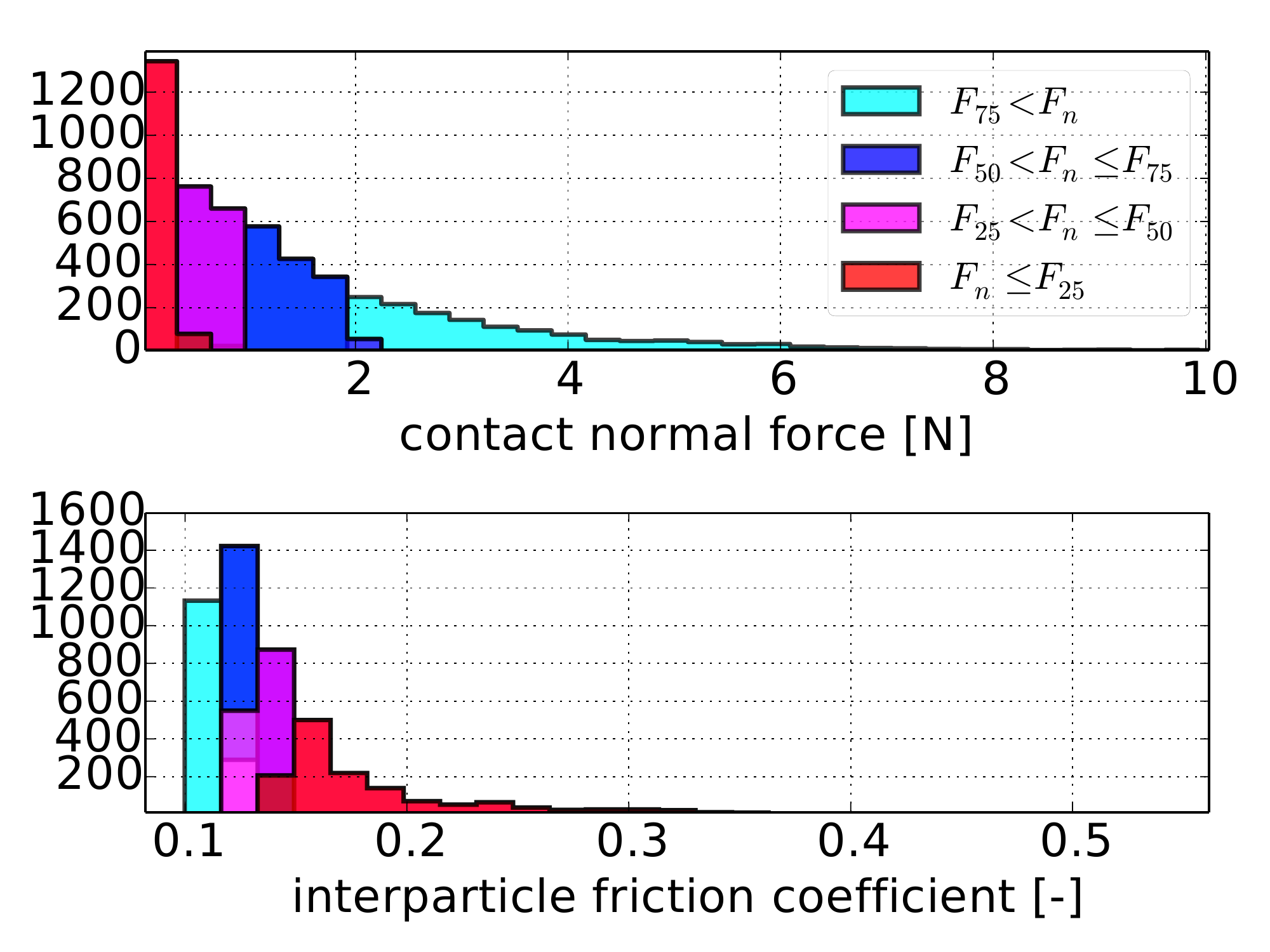}
\caption{Upper plot: Histograms of normal forces,  $F_n$, sorted ascendantly, divided into four equally sized groups. Lower plot: Histograms of corresponding $\mu$ values. %
}\label{fig:FnMuHist}
}
\end{figure}

For an investigation of the distribution of normal forces and interparticle friction for particle-particle contacts, the final state of a simulation with paired spheres and  $\sigma_n=$~24.2~kPa is considered. 
 For the presentation in Figure~\ref{fig:FnMuHist} the normal forces are sorted ascendantly and then divided in four equally sized groups (between the quartiles $F_{25}, F_{50}, F_{75}, F_{100}$). For each group, a histogram of the normal force is drawn in the upper part of the Figure.  In the lower part the histograms of the $\mu$ values belonging to each group are displayed. It can be seen that  25\% of all contacts transfer nearly no normal force. With growing magnitude of the normal force, its occurrence decays fast. The maximal observed force is about 16~N. The lowest normal forces result in the highest values of interparticle friction, where the upper bound of the interparticle friction is given by $\mu_0 + c_1 = 0.922$. In the considered settings all spheres are of equal size and material, therefore the equivalent contact radius and equivalent Young modulus  are the same for all contacts. Thus, an increase in normal force results directly in  a decrease of interparticle friction. The lowest value for $\mu$ observed in this simulation is 0.1. 

\section{Conclusions}
This work deals with the stress dependency of the interparticle friction coefficient and the resulting stress dependency in bulk friction.   In  \cite{Haertl2008} and \cite{Haertl2011}, H\"artl and Ooi,  conducted direct shear tests with single or paired glass beads.  For single spheres, the bulk friction was nearly constant, as it is frequently reported in literature. For paired spheres, a stress dependency in the bulk friction was observed, i.e.,~with increasing applied  normal stress the bulk friction decreased.  DEM simulation of the shear tests using Coulomb's law with a constant coefficient of friction for interparticle contacts, representing the state of the art, were conducted in \cite{Haertl2008} and \cite{Haertl2011} as well as in the current work. It turned out that neither for simulations with single spheres nor in simulations with paired sphere a stress dependency in the bulk friction could be seen. While the obtained DEM results are in good accordance with the experimental data for single spheres, the DEM results for paired spheres lack the stress dependence and are thus do not match experimental results. 

In tribology, it is known that the friction coefficient in a contact is not constant, as it is assumed in Coulomb's law, but depends on several factors, such as contact normal load, relative motion, surface roughness, contact temperature and contact conditions (dry, wet, lubricated, \ldots), etc.   
 It is assumed by the authors that the observed normal stress dependency of the bulk friction coefficient is caused by such tribological effects.  Thus, a more tribological tangential contact law is implemented in DEM, where the interparticle friction depends on the current normal stress in the contact. This approach aims at the improvement of the prediction quality with respect to the observed normal stress dependency of the bulk friction coefficient. 
 The proposed model is supported by measurements from literature, where qualitatively a decay of interparticle friction coefficient with increasing normal stress is found for several materials.   
    As it is future work, to obtain detailed measurement data of interparticle friction of glass over the complete range of normal stresses, 
     for now the stress dependency is derived from measurements of the bulk behaviour. Therefore, at first a parameter  study is conducted followed by the calibration using  experimental data. It turns out that only two of the three parameters need to be calibrated while one can be set constant. 
 The simulation results, obtained with the stress dependent model, are then compared to those, obtained with constant interparticle friction, and the experimental results.   
  The effect of the proposed model is qualitatively similar in simulations for single spheres as well as for paired spheres. Quantitatively, the magnitude of the stress dependency in the bulk behaviour is different for both particle shapes. %
 For single spheres, both the constant and the stress dependent results agree well with the measurements, where the stress dependent results tend to give a bulk friction, which is slightly too low. For the paired spheres the constant interparticle friction give poor results, while the stress dependent model shows good accordance. With these findings it can be concluded that the stress dependency of bulk friction coefficient, observed in the experiments, can be explained with a stress dependent interparticle friction model.

\section*{Acknowledgments}
\noindent
The authors gratefully acknowledge funding of the  Austrian Science Fund (FWF) for the project P 27147-N30: Short- and Long-Term Behaviour of Solid-Like Granular Materials.

This work was accomplished at the VIRTUAL VEHICLE Research Center in Graz, Austria. The authors would like to acknowledge the financial support of the COMET K2 - Competence Centers for Excellent Technologies Programme of the Austrian Federal Ministry for Transport, Innovation and Technology (bmvit), the Austrian Federal Ministry of Science, Research and Economy (bmwfw), the Austrian Research Promotion Agency (FFG), the Province of Styria and the Styrian Business Promotion Agency (SFG).

\section*{References}

\newpage

\clearpage

\end{document}